\documentclass{iopart}

\usepackage{epsfig}
\usepackage{amscd,amssymb}
\usepackage{graphicx,xspace}
\usepackage{pstricks}
\usepackage[left=3cm,top=3cm,right=3cm,bottom=3cm]{geometry}
\usepackage{color}

\begin{document}

\newcommand{\atanh}
{\operatorname{atanh}}
\newcommand{\ArcTan}
{\operatorname{ArcTan}}
\newcommand{\ArcCoth}
{\operatorname{ArcCoth}}
\newcommand{\Erf}
{\operatorname{Erf}}
\newcommand{\Erfi}
{\operatorname{Erfi}}
\newcommand{\Ei}
{\operatorname{Ei}}
\newcommand{\Pcal}
{{\mathcal P}}
\newcommand{\Prm}
{{\mathrm P}}

\title[EVS from RSRG : Brownian Motion, Bessel Processes and Continuous Time
  Random Walks]{Extreme value statistics from the Real Space
  Renormalization Group : Brownian Motion, Bessel Processes and Continuous Time Random Walks}

%\titlerunning{Field theory conjecture for loop-erased random walks}        % if too long for running head

\author{Gr{\'e}gory Schehr}

\address{Laboratoire de Physique Th\'eorique, 
Universit\'e Paris-Sud, 91405 Orsay Cedex, France }

\author{Pierre Le Doussal}

\address{CNRS-Laboratoire de Physique Th\'eorique de l'Ecole Normale Sup\'erieure, 
24 rue Lhomond, 75231 Paris Cedex-France
%\thanks{LPTENS is a Unit\'e Propre du C.N.R.S.
%associ\'ee \`a l'Ecole Normale Sup\'erieure et \`a l'Universit\'e Paris Sud}
}

\date{Received:  / Accepted:  / Published }

\begin{abstract}
We use the Real Space Renormalization Group (RSRG) method to study
extreme value statistics for a variety of Brownian motions, free or
constrained such as the Brownian bridge, excursion, meander and
reflected bridge, recovering some standard results, and 
extending others. We apply the same method to compute the distribution
of extrema of Bessel processes. We briefly show how the continuous
time random walk (CTRW) corresponds to a non standard fixed point of the RSRG
transformation.  
\end{abstract}

\maketitle

\section{Introduction}

There is recent a regain of interest for extreme value statistics (EVS) and
its applications to physics \cite{bm, dean_majumdar, pld_carpentier,
  racz, bertin_clusel} biology \cite{katz,malley} and finance
\cite{embrecht, bouchaud_satya}. Characterizing the statistical
properties of the maximum $X_{\max}$ (or the minimum $X_{\min}$) of a
set of $N$ random variables $X_1, X_2, \cdots, X_N$ has recently found
many applications in these various areas of research, particularly in
statistical physics. Indeed, being at the heart of
optimization problems, EVS plays a major role in the study of
disordered and glassy systems. 

The EVS of independent, or weakly correlated, and identically
distributed random variables is now well understood, thanks to the
identification, in the large $N$ (thermodynamical) limit, of three
distinct universality classes depending on the parent distribution of
the $X_i$'s \cite{gumbel}. The difficult cases arise when the random
variables are strongly correlated, which happens to be the case of
interest in many problems encountered in statistical physics. This is
the case for instance of the directed 
polymer in random media and its connections to Kardar-Parisi-Zhang 
equation \cite{kpz} or the statistical physics of a single
particle in a logarithmically correlated random potential
\cite{pld_carpentier}, closely related to Derrida's Random Energy
Model \cite{derrida_rem}.          

The most important example of a set of strongly correlated random
variables is of course Brownian motion (BM). And despite the fact that it has been 
widely studied \cite{feller, revuz, borodin}, there has been a
renewed interest in the study of functionals involving Brownian motions \cite{yor_functional,
  satya_functionals}, and in particular in the context of extreme
statistics \cite{satya_airy, schehr_airy, comtet_max_rw, satya_yor, rambeau_airycrossover}. Considering a Brownian
motion $u(x)$ on the interval $x \in [0,L]$, subject to $u(0)=0$, it reaches its maximum $u_m$ at time $x=x_m$. The basic question we ask is : what is the joint distribution $P_L(u_m,x_m)$ of these random variables? For instance, for unconstrained Brownian motion, the (marginal) distribution ${\cal P}_L(x_m)$ is given by ${\cal P}_L(x_m) = \frac{1}{\pi}x_m^{-1/2} (L-x_m)^{-1/2}$ or equivalently the cumulative distribution ${\rm Proba}(x_m \leq x) = \frac{2}{\pi} \sin^{-1}(\sqrt{x/L})$, which is the classical L\'evy's arcsine law \cite{levy_arcsine}. These classical results for BM for $P_L(u_m,x_m)$ have been recently obtained using Feynman-Kac formula and further extended to a wider class of constrained Brownian motions, including Brownian excursions, meanders and reflected Brownian bridge \cite{satya_yor}. Interestingly, it was shown that this joint distribution $P_L(u_m,x_m)$ for a Brownian motion and Brownian bridge arises naturally in the study of the convex hull of planar Brownian motions \cite{satya_convex}. Recently, such results for the distribution of the maximum ${\rm P}_L(u_m)$ have been obtained, using also a path-integral formalism, for the case of multidimensional processes, where one considers $p$ non
intersecting Brownian walkers (vicious walkers)~\cite{schehr_watermelons, nadal_watermelons}.  

On the other hand, it is interesting to extend these studies to other stochastic processes. A natural one is the Bessel process, which is the radius of $d-$dimensional Brownian motion. The study of the distribution of the maximum ${\rm P}_L(u_m)$ for a Bessel bridge has actually been widely studied in probability theory and statistics, in particular in the context of the generalization
of the Kolmogorov-Smirnov test (corresponding to $d=1$)
\cite{kolmogorov_smirnov}. The marginal distribution ${\rm P}_L(u_m)$
for such Bessel bridges was first computed by Gikhman and Kiefer
\cite{gikhman, kiefer} and then generalized to any real $d$ by Pitman
and Yor \cite{pitman_yor}, while much less is known about ${\cal
  P}_L(x_m)$. In Ref. \cite{gikhman, kiefer,pitman_yor}, these results
are obtained using rather complicated tools from probability theory,
including the so called ``agreement formulae'' which, albeit rigorous,
are not very physically intuitive (see however
Ref. \cite{satya_yor}). To our knowledge, the extreme statistics of
the Bessel process has not been studied using the tools of statistical
physics.

In this paper we will use the Real Space Renormalization Group (RSRG)
to study these problems of extreme value statistics. The RSRG was
first devised for the study of quantum spin chains with disorder
\cite{fisher1,igloimonthus,fisheryoung,monthusFS}. It was then applied
\cite{sinaiRSRG} to the study of the Sinai model \cite{sinai},
{\it i.e.} the Arrhenius motion (diffusion) of a particle in a random
potential $u(x)$ which is itself a random walk, {\it i.e.} it becomes a one
dimensional Brownian motion in the scaling limit. In both cases the
idea is to decimate iteratively the smallest energy levels (for
quantum systems) or barriers (for Arrhenius dynamics), defined as
variables $F=|u(x_i)-u(x_{i+1})|$ where $(x_i,x_{i+1})$ are pairs of
successive of local min-max. At a given stage, indexed by the
"decimation scale" called $\Gamma$, the landscape does not contain any
barrier smaller than $\Gamma$. While it is only asymptotically exact
for the above mentioned problems, it was later realized
\cite{pldmonthus} that it is also a fully exact and quite powerful
tool to study extreme value statistics, and that it leads to solvable
equations when applied to a broad class of processes, which includes,
but is not restricted to, the Brownian motion. In fact it gives much
more information than the frequently asked questions in extreme value
statistics as it also allows one to compute the statistics of $h$-extrema
(as introduced in the mathematical literature by Neveu and Pitman
\cite{neveu_hextrema}), {\it
i.e.} the extrema separated by a barrier larger than
$h$, and here $h \equiv \Gamma$. In Sinai's walk,
these are "important extrema" of the energy 
landscape, which are separated by
large enough barriers such that the process is not authorized to
backtrack on more than $\Gamma$ (see Ref. \cite{bovier_hextrema} for $h$-extrema in the context of Sinai's walk). In Ref. \cite{pldmonthus} the RSRG was used 
to obtain the extremal statistics of the toy model, {\it i.e.} Brownian
motion plus a quadratic well. 

The aim of this paper is to explain and illustrate how the RSRG allows one 
to recover known results about extrema of the most standard
constrained Brownian motions, {\it i.e.} the bridge, the meander, the
reflected BM and the excursion. It is in a sense an overkill, as we will
see the method is too powerful and we only use it in the limit $\Gamma
\to \infty$. However, in view of the versatility and importance of
this method, we think that it is useful to establish some links with
more classical approaches and results. Since we want to be mostly
pedagogical, we restrict to symmetric BM, the case of a bias being a
straightforward extension. Note that even for the Brownian motion we
compute a rather complicated object, depicted in
Fig. \ref{fig_unconstrained}, at the same price as the standard, well
known result for the distribution of the maximum. Next we study the
Bessel process ({\it i.e.} the radius of a $d$-dimensional Brownian motion)
and obtain its extremal statistics. Finally we close by a short
Section on the extensions to continuous time random walks (CTRW). Some of our results are standard well known results, derived here in a rather different fashion, others are extensions and likely to be not so well known. 

\section{Extreme value statistics of some classical examples of constrained Brownian motion}

\subsection{Short review of RSRG}

\begin{figure}[t]
  \centering
  \includegraphics[width=0.9\linewidth]{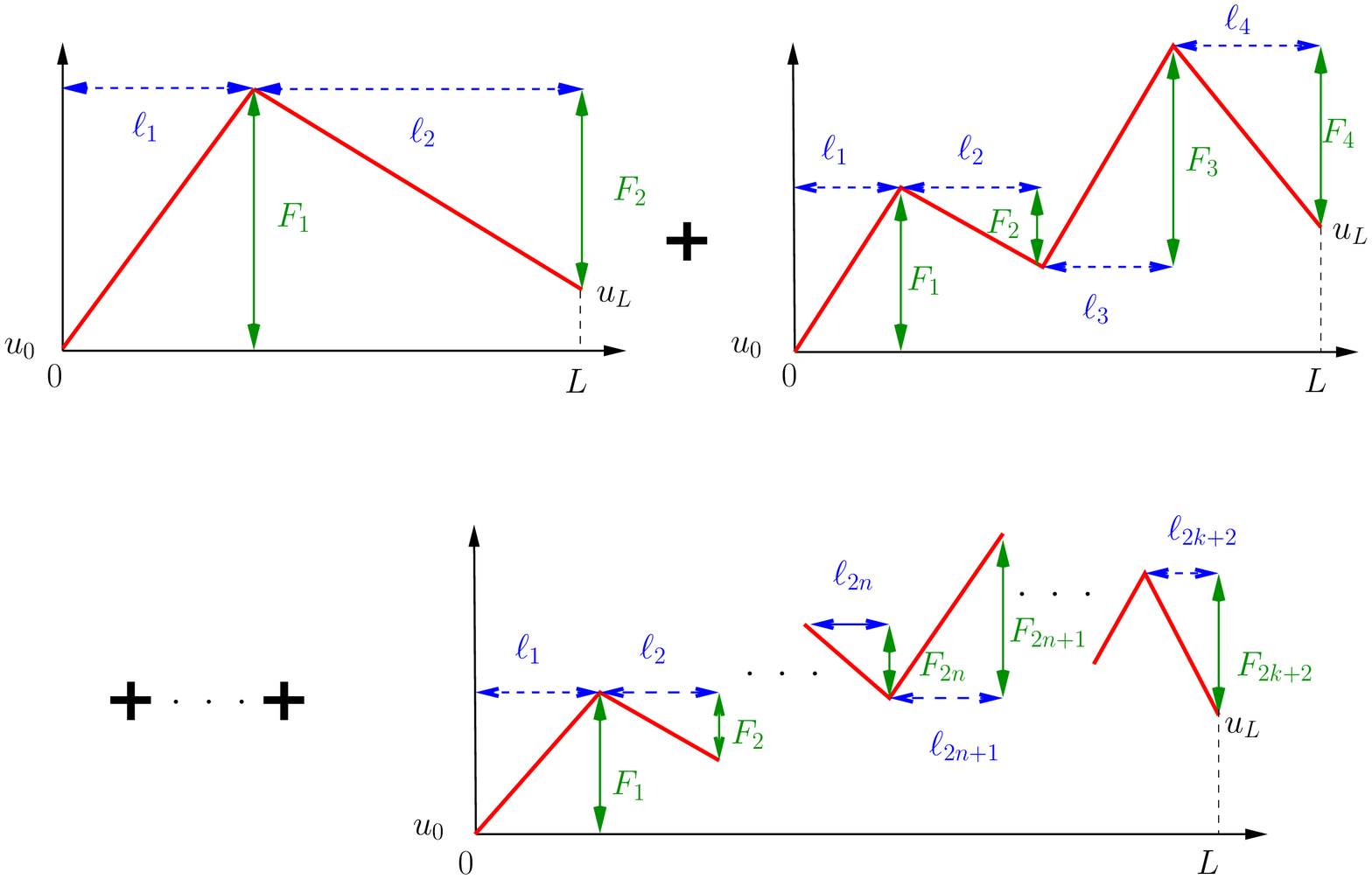}
 \caption{Schematic representation of the finite size RSRG measure
 (\ref{fs}) for the renormalized landscape (on the $x$ axis, the
 $\ell_j$'s are the bond length, and on the vertical $y$ axis the $F_j$'s
 are the barriers). It is a sum of 
 terms with $k=1,2,..$ valleys ({\it i.e.} sets of two consecutive bonds) not yet decimated at scale $\Gamma$, {\it i.e.} of barriers larger than $\Gamma$ (in vertical $y$ direction). The edge bonds (first and last) cannot be decimated, only the bulk bonds (all the others) are.}
    \label{fig:rsrg0}
\end{figure}

Let us briefly recall the main results of the RSRG method. Here we restrict to the Brownian motion. The general case, introduced in Ref. \cite{pldmonthus}, is recalled later. The first statement is that the probability measure of the symmetric Brownian motion $u(x)$ with $x \in [0,L]$ and $u(0)=u_0$, $u(L)=u_L$, can be written as the following sum:
\begin{eqnarray} \label{fs} 
\fl \bar l_\Gamma E_\Gamma(F_1,\ell_1)E_\Gamma(F_2,\ell_2) \delta(L-(\ell_1+\ell_2))
\delta(u_L-(u_0+F_1-F_2)) \nonumber \\
\fl + \sum_{k=1}^{\infty} 
 \bar l_\Gamma  E_\Gamma(F_1,\ell_1)  
\prod_{j=2}^{2k+1} P(F_{j},\ell_{j})   E_\Gamma(F_{2k+2},\ell_{2k+2} )  \delta(L- \sum_{i=1}^{2 k+2} \ell_i) \delta(u_L-(u_0+\sum_{j=1}^{2k+2} (-1)^{j+1} F_{j})) \;.
\end{eqnarray}
where $\bar l_\Gamma=\Gamma^2$.
In the (finite size) RSRG language \cite{sinaiRSRG} this is called the renormalized landscape at scale $\Gamma$,  with bond lengths $\ell_i=x_i-x_{i-1}$ and barriers $F_{2 n}= u(x_{2 n-1})-u(x_{2n})$ or $F_{2 n+1} = u(x_{2 n+1})-u(x_{2 n})$ where $x_{2n}$ are local minima and $x_{2n+1}$ are local maxima, and it holds for a large class of random walk landscape with Markov property. It is illustrated in Fig. \ref{fig:rsrg0}. 
The transformation which increases $\Gamma$ is called the
{\it decimation}. It iteratively reduces the list of extrema $x_0=0<x_1<...<
x_{2k+2}<x_L=L$ by removing a pair of adjacent min-max (or max-min)
from the list corresponding to the smallest $F$ in the system, when
$\Gamma$ crosses that value, in effect gluing three bonds into
one. This results in a RG ({\it i.e.} evolution) equation for
$P_\Gamma(F,\ell)$ (bulk bonds) and $E_\Gamma(F,\ell)$ (edge bonds)
which eventually reaches a fixed point form which, under some
conditions, corresponds exactly to the Brownian motion. The edge bonds
play a special role as they cannot be decimated, they can only grow by
decimation of their nearest neighbor. The total measure (\ref{fs}) is
a sum of measures for the events where there remain $2 k+2$ bonds in
the system, {\it i.e.} $k$ valleys of bulk bonds and two edge bonds, with
$k=0,1,..$. At large $\Gamma$, for $L$ finite, there remains only the
term $k=0$ which plays a special role. The product form reflects the
Markovian nature of the landscape, which is preserved by decimation,
hence is either an exact consequence of the choice of a Markovian
initial landscape, or a consequence of the convergence to the fixed
point landscape. The only constraint is the fixed total length,
implemented by the delta functions. The factor $l_\Gamma = \Gamma^2$ is nothing but the average bond length which ensures the normalization of the total probability to unity. The fixed point form for the bond length probability, {\it i.e.} the one corresponding to the BM, takes in 
Laplace $P_\Gamma(F,p)=\int_0^L e^{-p \ell} P_\Gamma(F,\ell)$, the simple form:
\begin{eqnarray}
&& P_\Gamma(F,p) = \frac{\sqrt{p}}{\sinh(\Gamma \sqrt{p})} e^{- (F - \Gamma) \sqrt{p} \coth(\Gamma \sqrt{p})} ~ \theta(F-\Gamma) \;, \label{bulk} \\
&& E_\Gamma(F,p) = \Gamma^{-1} e^{- F \sqrt{p} \coth(\Gamma \sqrt{p})}  \label{eg} \;.
\end{eqnarray}
As mentioned above $P_\Gamma(F,\ell)$ can be related to a sum over all BM which are constrained to climb from $0$ to $F$ in a time $\ell$ while remaining in the interval $]0,F[$ and with no descent of more than $\Gamma$. The RSRG equations which yield the fixed point forms (\ref{bulk}, \ref{eg}) are recalled in Appendix A. 

\subsection{Free Brownian and Brownian Bridge}

In all cases one considers a Brownian motion $u_x$ on the interval
$[0,L]$, with $\langle du_x^2 \rangle = 2 dx$, with initial condition
$u_0=0$. Note that this corresponds to a diffusion coefficient $D=1$,
in contrast to the value $D=1/2$ used in Ref. \cite{satya_yor} to which we will
refer in the following. 
\begin{figure}[h]
\centering
\includegraphics[width=0.6\linewidth]{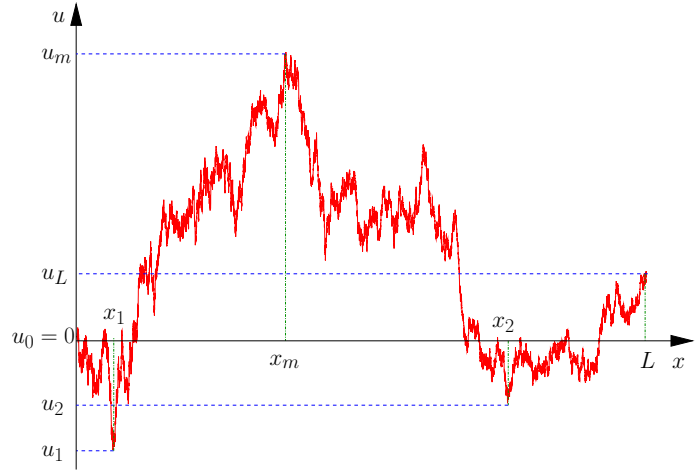}
\caption{Unconstrained Brownian motion $u_x$ starting at $u_0=0$ and
  ending at $u_L$ which is free. It reaches its maximum $u_m$ at $x_m$ and has a
  minimum $u_1$, reached at $x_1$, on the segment $[0,x_m]$ and a
  minimum $u_2$, reached at $x_2$ on the segment $[x_m,L]$.}\label{fig_unconstrained} 
\end{figure}
While the free Brownian is unconstrained at $x=L$ (see Fig. \ref{fig_unconstrained}), the
Brownian Bridge is defined by the constraint $u_L=0$ (see
Fig. \ref{fig_constrained} a)). It is natural to consider the joint probability density $P_L(u_{m},x_{m},u_L)$
that the free Brownian ends at $u_L$ and has maximum value $u_{m}$ at
position $x_{m}$. From the above considerations it is obtained from
RSRG by considering the last block (see
Fig. \ref{fig_lastrsrgblock}), and is given by: 
\begin{eqnarray}\label{last_block}
&& P_L(u_{m},x_{m},u_L) = \lim_{\Gamma \to \infty} 
\Gamma^2 E_\Gamma(F_1,x_m) E_\Gamma(F_2,L-x_m) \;,
\end{eqnarray}
with $F_1=u_{m}$ and
$F_2=u_{m}-u_L$. Using the result (\ref{eg}) for the
edge bond, 
%\begin{eqnarray}
%&&E_\Gamma(F,p) =  \int_0^\infty dl E_\Gamma(F,l) e^{-p l} = \Gamma^{-1} e^{- F% \sqrt{p} \coth(\Gamma \sqrt{p})} \;,
%\end{eqnarray}
one finds, in the large $\Gamma$ limit:
\begin{eqnarray}
\lim_{\Gamma \to \infty} \Gamma E_\Gamma(F,p) = e^{- F \sqrt{p}}
  \;, 
\end{eqnarray}
which gives after inverse Laplace transformation :
\begin{eqnarray}\label{edge_brownian}
\lim_{\Gamma \to \infty} \Gamma E_\Gamma(F,l) = \frac{F}{2
    \sqrt{\pi} l^{3/2}}  
e^{-F^2/(4 l)} \;.
\end{eqnarray}
\begin{figure}[h]
\centering
\includegraphics[width=0.8\linewidth]{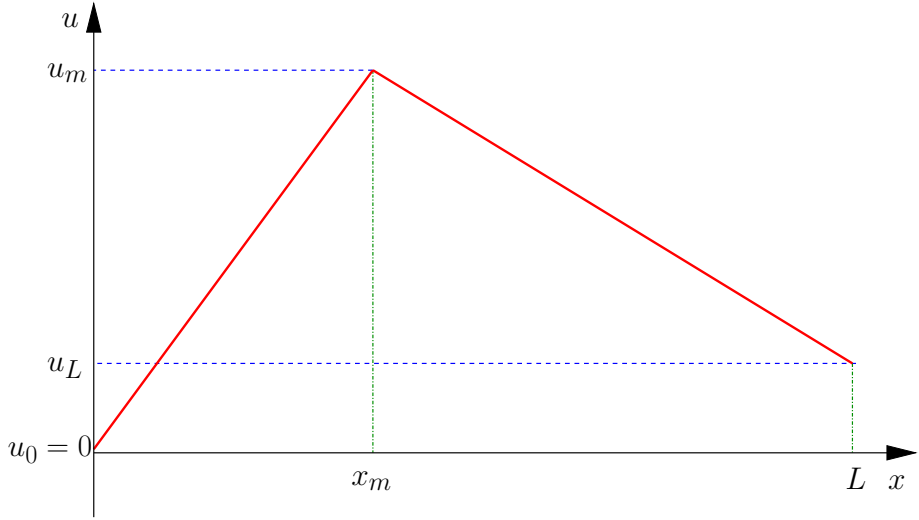}
\caption{ Schematic representation of the renormalized landscape
  associated to the one in Fig. (\ref{fig_unconstrained}) after
  the last RSRG step, {\it i.e.} the last decimation. The measure of
  this last block gives the 
  joint distribution $P(x_m,u_m,u_L)$ (see
  Eq. (\ref{last_block})).}\label{fig_lastrsrgblock}
\end{figure}

Thus using Eq. (\ref{last_block}) together with the expression for the
edge bond in Eq. (\ref{edge_brownian}) one has
\begin{eqnarray}\label{joint_brownian_unconst}
&& P_L(u_{m},x_{m},u_L)  = \frac{1}{4 \pi} \frac{u_{m} (u_{m}-u_L)}{x_{m}^{3/2} (L-x_{m})^{3/2}}
e^{-\frac{u_{m}^2}{4 x_m} -\frac{(u_{m}-u_L)^2}{4 (L-x_m)}} \theta(u_m-u_L) \theta(u_m) \;.
\end{eqnarray}
Integration over $x_m$ and $u_m$ yields the propagator $W(u_L,L)$
\begin{eqnarray}\label{propag_bm}
&& W(u_L,L) = \int^{\infty}_{\max{(0,u_L)}} du_{m} \int_0^L dx_m P_L(u_{m},x_m,u_L)
  = \frac{1}{\sqrt{4 \pi L}} e^{- \frac{u_L^2}{4 L}} \;. 
\end{eqnarray}

We now obtain the result for the free Brownian motion by integrating the
formula (\ref{joint_brownian_unconst}) over the final position $u_L$:
\begin{eqnarray}
P^{\rm free}_L(u_{m},x_{m}) = \int_{-\infty}^{u_m} du_L
  P_L(u_{m},x_m,u_L) = \theta(u_m) \frac{u_m}{2 \pi \sqrt{L-x_m} x_m^{3/2}}
  e^{-  \frac{u_{m}^2}{4 x_m}}  \;.
\end{eqnarray}
This yields the well known results for the marginal distribution of the position of the maximum
$x_m$ for the free Brownian motion
\begin{eqnarray}
&& \Pcal^{\rm free}_L(x_{m})  = \frac{1}{L} \tilde \Pcal^{\rm
    free}\left(\frac{x_m}{L} \right) \;, \; \tilde \Pcal^{\rm
    free} (x) = \frac{1}{ \pi \sqrt{x(1-x)}}  \;,
\end{eqnarray}
which is the well known L\'evy's ``arcsine law'' \cite{levy_arcsine}, as well as the distribution of the maximum $u_m$
\begin{eqnarray}
\Prm^{\rm free}_L(u_{m})  = \frac{1}{L^{1/2}} \tilde \Prm^{\rm
    free}\left(\frac{u_m}{L^{1/2}} \right) \;, \; \tilde \Prm^{\rm
    free} (x) = \theta(x)  \frac{e^{- \frac{x^2}{4}}}{\sqrt{\pi}} \;,
\end{eqnarray}
where $\theta(x)$ is the Heaviside step function, $\theta(x) = 0$ if $x < 0$ and $\theta(x) = 1$ if $x \geq 0$. 

From Eq. (\ref{joint_brownian_unconst}) we also immediately get the
result for the Brownian Bridge, where $u_L = 0$ (see
Fig. \ref{fig_constrained} a)): 
\begin{eqnarray}\label{joint_bridge}
\fl P^{\rm bridge}_L(u_{m},x_{m})  =
  \frac{P_L(u_m,x_m,u_L=0)}{\int^{\infty}_{0} du_{m}
    \int_0^L dx_m P_L(u_{m},x_m,u_L=0)} = \theta(u_m)\sqrt{\frac{L}{4\pi}}\frac{u_{m}^2}{x_{m}^{3/2} (L-x_{m})^{3/2}}
e^{-\frac{ L u_{m}^2}{4 x_m (L-x_m)} }  \;. \nonumber  \\
&&
\end{eqnarray}
This yields the marginal distributions of $x_m$ on the one hand and of
$u_m$ on the other hand
\begin{eqnarray}
&& \Pcal^{\rm bridge}_L(x_{m})  = \frac{1}{L} \;,\\
&& \Prm^{\rm bridge}_L(u_{m})  = \frac{1}{L^{1/2}} \tilde \Prm^{\rm
    bridge}\left(\frac{u_m}{L^{1/2}} \right) \;, \; \tilde \Prm^{\rm
    bridge} (x) = 2 \theta(x) x e^{-x^2} \;.
\end{eqnarray}
We thus see that the RSRG method allows to recover standard results
for the relatively easier cases of free Brownian motion and the Brownian
bridge. Let us now investigate more complicated instances of other
classical constrained Brownian motions.

{\bf Notational remark} : in the following we will use invariably the notation $P_L(\cdots)$, independently of the number of its arguments,  
to denote some joint distribution. Similarly, we will also use the same notation to denote its Laplace transform with respect of one or several of its arguments. The choice of the notations for the different variables should be clear enough to avoid any confusion.

\subsection{Reflected Brownian motion, Brownian excursion and meander}

\begin{figure}
\centering
\includegraphics[width=0.8\linewidth]{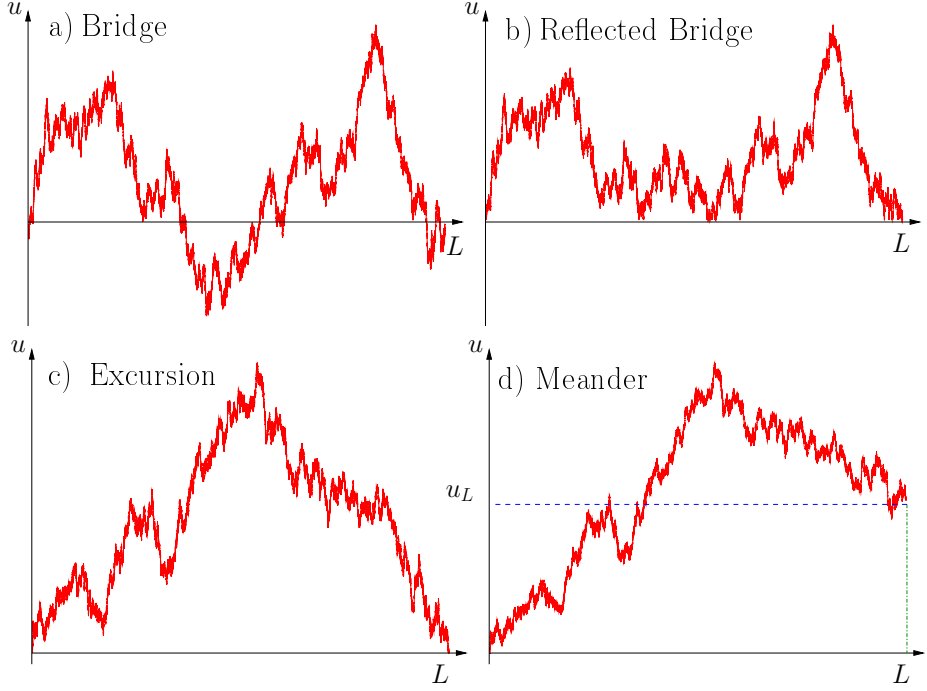}
\caption{Various constrained Brownian motions considered in this paper : a) Brownian bridge which is a Brownian motion constrained to start and end in $0$, {\it i.e.}
which $u_0 = u_L = 0$. b) Reflected Brownian bridge, {\it} $u_x = |u^0_x|$ where $u^0_x$ is the standard Brownian motion, and constrained to  start and end in $0$. c) Brownian excursion, {\it i.e.} Brownian motion constrained to start and end in $0$, remaining positive in between. d) Brownian meander, {\it} Brownian motion starting in $0$, ending anywhere in $u_L > 0$ and remaining positive in between.}\label{fig_constrained}
\end{figure}

In this section, we study the cases of the reflected Brownian motion,
excursion and meander, see 
Fig.~\ref{fig_constrained} b), c) and d) respectively. The joint
distribution $P_L(u_m,x_m)$ of the maximum $u_m$ and its position
$x_m$ for these three cases was recently computed using both path integral
techniques as well as the so called rigorous ``agreement formulae'' in
Ref. \cite{satya_yor}. Using RSRG, all these cases can be obtained
from the probability $P_L(u_1,x_1,u_m,x_m,u_2,x_2,u_L)$ that an
unconstrained Brownian $u_x$ starting at $u_0=0$ ends at $u_L$, reaches its maximum $u_m$ at $x_m$ and that its minimum $u_1$ on the segment $[0,x_m]$ is
reached at $x_1$, and that its minimum $u_2$ on the segment $[x_m,L]$
is reached at $x_2$ (see Fig. \ref{fig_unconstrained}). It can be
obtained from the edge bond probabilities (including information on
the minima) as explained in \ref{appendix_edge_max} as: 
\begin{eqnarray}
&& \!\!\!\!\!\!\!\!\!\!\!\! \!\!\!\!\!\!\!\!\!\!\!\!  \!\!\!\!\!\!\!\!\!\!\!\! P_L(u_1,x_1,u_m,x_m,u_2,x_2,u_L) = E(u_m,x_m,-u_1,x_1)  E(u_m-u_L,L-x_m,-u_2+u_L,L-x_2) \;, \label{block_struct}\\
&& \!\!\!\!\!\!\!\!\!\!\!\!  E(u_m,x_m,u,x) =  \lim_{\Gamma \to \infty} \Gamma E_\Gamma(u_m,x_m,u,x) \nonumber \;.
\end{eqnarray}
We show in \ref{appendix_edge_max} that in Laplace variables:
\begin{eqnarray}
&& E(u_m,p,u,q) =  \int_0^\infty dx_m \int_0^\infty dx e^{- p x_m -q x} E(u_m,x_m,u,x)  \nonumber \;, \\
&& E(u_m,p,u,q) = \frac{ \sinh(\sqrt{p+q} u_m)}{\sinh(\sqrt{p+q} (u+u_m)) } \frac{\sqrt{p}}{\sinh(\sqrt{p} (u+u_m))} \label{E_laplace}  \;.
\end{eqnarray}
Performing Laplace inversions (see Eq. (\ref{inverselaplace_1}, \ref{inverselaplace_2})) one finds:
\begin{eqnarray}
&& E(u_m,x_m,u,x) =  \sum_{n,m=0}^\infty 4 \pi^3 n m^2 (-1)^{n+m}
\frac{\sin(\frac{\pi u_m}{u+u_m} n)}{(u_m+u)^5} e^{- \frac{\pi^2}{(u+u_m)^2} (n^2 x + m^2 (x_m-x))} \label{E_direct} \;.
\end{eqnarray}
Notice the scaling forms, directly read from Eqs. (\ref{E_laplace}, \ref{E_direct})
\begin{eqnarray}
&& \!\!\!\!\!\!\!\!\!\!\!\! E(u_m,x_m,u,x) = \frac{1}{u_m^5} \tilde E\left(\frac{x_m} {u_m^2}, \frac{u}{u_m},\frac{x}{u_m^2}\right) \label{scaling_direct} \;,\\
&& \!\!\!\!\!\!\!\!\!\!\!\! E(u_m,p,u,q) = \frac{1}{u_m} \tilde E
  \left(p u_m^2,\frac{u}{u_m},q u_m^2\right) \;, \; \nonumber \\
{\rm with} \; &&\tilde E(\tilde p,\tilde u,\tilde q) = \int_0^\infty
d\tilde x_m \int_0^\infty d \tilde x e^{-\tilde p \tilde x_m-\tilde q
  \tilde x} \tilde E(\tilde x_m, \tilde u, \tilde x) \nonumber \;. \label{scaling_laplace}
\end{eqnarray}
From Eqs. (\ref{block_struct}, \ref{E_direct}) one gets the probability density of the event depicted in Fig. \ref{fig_unconstrained}:
\begin{eqnarray}\label{full_expr}
&& \fl P_L(u_1,x_1,u_m,x_m,u_2,x_2,u_L) = \theta(u_L- u_2) \theta(u_m-u_L)
 \sum_{n_1,m_1,n_2,m_2=0}^\infty  16 \pi^6 n_1 n_2 m_1^2 m_2^2 (-1)^{n_1+m_1+n_2+m_2} \nonumber \\
&& \fl \times \frac{\sin(\frac{\pi u_m}{u_m-u_1} n_1) \sin(\frac{\pi (u_m-u_L)}{u_m-u_2} n_2) }{(u_m-u_1)^5 (u_m-u_2)^5} 
e^{- \frac{\pi^2}{(u_m-u_1)^2} (n_1^2 x_1 + m_1^2 (x_m-x_1)) - \frac{\pi^2}{(u_m-u_2)^2} (n_2^2 (L-x_2) + m_2^2 (x_2-x_m))  }  \;.
\end{eqnarray}
From this object (\ref{full_expr}) we now obtain the probability density $P_L(u_m,x_m,u,x,u_L)$ that an unconstrained Brownian $u_x$ starting at $u_0=0$ ends at $u_L$, reaches its maximum $u_m$ at $x_m$ and its minimum $u$ at $x$. It is given by
\begin{eqnarray}\label{min_max}
&& P_L(u_m,x_m,u,x,u_L) = \theta(x_m-x) \int_{u}^{u_L} du_2
  \int_{x_m}^L dx_2 P_L(u,x,u_m,x_m,u_2,x_2,u_L) \\ 
&& + \theta(x-x_m)\int_{u}^{0} du_1  \int_{0}^{x_m} dx_1
P_L(u_1,x_1,u_m,x_m,u,x,u_L)  \;.
\end{eqnarray}
After some algebra, left in \ref{appendix_minmax} one obtains
\begin{eqnarray}\label{min_max_explicit}
&&\fl P_L(u_m,x_m,u,x,u_L) = \theta(x_m-x) \sum_{n_1,m_1}^\infty  4 \pi^3 n_1 m_1^2 
  (-1)^{n_1+m_1} \frac{\sin(\frac{\pi u_m}{u_m-u} n_1)}{(u_m-u)^5}
  e^{- \frac{\pi^2}{(u_m-u)^2} (n_1^2 x + m_1^2 (x_m-x))} \nonumber  \\   
&& \times  \sum_{n_2=0}^\infty 2 \pi (-1)^{n_2+1} n_2  \frac{\sin(\frac{\pi
  (u_L-u)}{u_m-u} n_2)}{(u_m-u)^2} e^{-
  \frac{n_2^2\pi^2}{(u_m-u)^2}(L-x_m)} \nonumber \\
&& + \theta(x-x_m) \sum_{n_2,m_2}^\infty  4 \pi^3 n_2 m_2^2 
  (-1)^{n_2+m_2} \frac{\sin(\frac{\pi u_m}{u_m-u} n_2)}{(u_m-u)^5}
  e^{- \frac{\pi^2}{(u_m-u)^2} (n_2^2 (L-x) + m_2^2 (x-x_m))} \nonumber \\  
&& \times \sum_{n_1=0}^\infty 2 \pi (-1)^{n_1+1} n_1  \frac{\sin(\frac{\pi
  (-u)}{u_m-u} n_1)}{(u_m-u)^2} e^{- \frac{n_1^2\pi^2}{(u_m-u)^2}x_m}   \;.
\end{eqnarray} 
In particular, one can check explicitly the following symmetry
\begin{eqnarray} \label{symmetry1}
P_L(u_m,x_m,u,x,u_L=0) = P_L(-u,x,-u_m,x_m,u_L=0) \;.
\end{eqnarray}

From the structure of Eq. (\ref{block_struct}) one obtains the useful expression for $\hat P_L(u_m,x_m,u_L,a)$ which is the joint probability of the maximum $u_m$ and of its position $x_m$ for a Brownian motion starting at $u_0 = 0$ and ending at $u_L$ and with a minimum value $u$ larger than $-a$. It reads :  
\begin{eqnarray}\label{useful_formula}
&& \!\!\!\!\!\!\!\!\!\!\!\!  \hat P_L(u_m,x_m,u_L,a) = \int_{-a}^0 du \int_0^L dx P_L(u_m,x_m,u,x,u_L) \nonumber \\
&& \!\!\!\!\!\!\!\!\!\!\!\!  =  \int_0^{u_L+a} du \int_0^{L-x_m} dx E(u_m-u_L,L-x_m,u,x) \int_0^a du' \int_0^{x_m} dx' E(u_m,x_m,u',x') \;.
\end{eqnarray}
\subsubsection{Reflected Brownian bridge}

The reflected Brownian is defined by $u_x=|u^0_x|$ where $u^0_x$ is
the usual Brownian, and initial condition $u_0=0$. We focus on the
reflected Brownian bridge, for which $u_L = 0$, see
Fig. \ref{fig_constrained} b). Therefore one has 
\begin{eqnarray}
&& P_L^{\rm{ref. \, bridge}}(u_m,x_m) = \sqrt{4 \pi L} \big(\int_{-u_{m}}^0
du \int_0^L dx P_L(u_{m}, x_{m},u,x,u_L=0) \\
&& + \int_0^{u_{m}} d\tilde u_m
\int_0^L d\tilde x_m P_L(\tilde u_m,\tilde x_m,-u_{m},x_{m},u_L=0) \big) \;,
\end{eqnarray}
where the prefactor $\sqrt{4 \pi L}$ comes from the fact we are
considering bridges (similarly to Eq. (\ref{joint_bridge}) above). Using the above symmetry (\ref{symmetry1}) one
sees that these two terms are actually identical, yielding
\begin{eqnarray}
P_L^{\rm ref. \, bridge}(u_{m},x_{m}) = 2\sqrt{4 \pi L} \int_{-u_{m}}^0
du \int_0^L dx P_L(u_{m}, x_{m},u,x,u_L=0) \;.
\end{eqnarray}
Therefore one has $P_L^{\rm ref. bridge}(u_{m},x_{m}) = 2\sqrt{4 \pi
  L} \hat P_L(u_m,x_m,u_L=0,u_m)$, and the above identity in Eq.~(\ref{useful_formula}) yields
\begin{eqnarray}
\fl  P_L^{\rm ref. \, bridge}(u_{m},x_{m}) = 2\sqrt{4 \pi L}  \int_0^{u_m} du \int_0^{L-x_m} dx E(u_m,L-x_m,u,x) \int_0^{u_m} du \int_0^{x_m} dx E(u_m,x_m,u,x) \;.\nonumber \\
&& 
\end{eqnarray}
From the scaling forms in Eq. (\ref{scaling_direct}) one obtains that
\begin{eqnarray}
P_L^{\rm ref. \, bridge}(u_{m},x_{m}) = 2\sqrt{4 \pi L} P_c\left( \frac{u_m}{x_m^2}\right)  P_c\left( \frac{L-u_m}{x_m^2}\right) \;,
\end{eqnarray}
where the Laplace transform of $P_c(z)$ is given by
\begin{eqnarray}
\!\!\!\!\!\!\!\!\!\!\!\!  \int_0^\infty dz e^{-p z} P_c(z) = \int_0^1 du \tilde E(p,u,q=0) =\int_0^1 du \frac{\sqrt{p}\sinh{(\sqrt{p})}}{(\sinh{(\sqrt{p}(u+1))})^2} = \frac{1}{2 \cosh{(\sqrt{p})}} \;.
\end{eqnarray}
By inverting the Laplace transform, one obtains
\begin{eqnarray}
P_c(z) = \pi \sum_{n=0}^\infty (-1)^n (n+\frac{1}{2}) e^{-\pi^2
  (n+1/2)^2 z} \;.
\end{eqnarray}
And finally, one has
\begin{eqnarray}\label{joint_reflected}
\fl P_L^{\rm ref. \, bridge}(u_{m},x_{m}) = 4 \pi^{5/2} L^{1/2}
\frac{1}{u_{m}^4} 
\sum_{m,n=0}^\infty (-1)^{n+m} (n+\frac{1}{2})(m+\frac{1}{2})  e^{-\pi^2
  (m+1/2)^2 \frac{x_{m}}{u_{m}^2}}  e^{-\pi^2
  (n+1/2)^2 \frac{L-x_{m}}{u_{m}^2}} \nonumber \;,\\
  && 
\end{eqnarray}
which gives back the formula obtained by Majumdar et al. in Ref. \cite{satya_yor} (see their
formula (28)). In particular, by integrating over $u_m$, one obtains the marginal distribution of $x_m$ as
\begin{eqnarray}
&&{\mathcal P}_L^{\rm ref. \, bridge}(x_m) = \frac{1}{L}\tilde
  {\mathcal P}^{\rm ref. \, bridge}\left(\frac{x_m}{L} \right)  \;,\\ 
&& \tilde {\mathcal P}^{\rm ref. \, bridge}(x) = 2 \sum_{n,m=0}^\infty (-1)^{m+n} \frac{(2m+1)(2n+1)}{\left[ (2n+1)^2 x + (2m+1)^2(1-x) \right]^{3/2}} \;.
\end{eqnarray}

\subsubsection{Brownian excursion}

Here one considers the case of Brownian excursions, constrained to start and end in $0$, remaining positive in between, see Fig. \ref{fig_constrained} c). 
In that case, one would naively impose that the minimum is zero
exactly and $u_L = 0$. However, one immediately sees on
Eq. (\ref{min_max}) that $P_L(u_m,x_m,u=0,x=0,u_L=0) = 0$. This is expected
since the Brownian motion has an infinite density of zero
crossings. Therefore, to compute the joint probability $P_L^{\rm ex}(u_m,
x_m)$ from $P_L(u_m,x_m,u,x,u_L=0)$ in Eq. (\ref{min_max}), one uses a
limiting procedure (see Ref. \cite{satya_airy, satya_yor} for a similar procedure used within Feynman-Kac formula): 
\begin{eqnarray}
P_L^{\rm ex}(u_m, x_m) = \lim_{\epsilon \to 0} \frac{\int_\epsilon^0 du \int_0^L dx P_L(u_m, x_m, u,x,u_L=0)}{\int_0^\infty du_m \int_0^L dx_m
  \int_\epsilon^0 du \int_0^L dx P_L(u_m, x_m, u,x,u_L=0)} \;.
\end{eqnarray} 
Following the same analysis as above, and using again the identity in Eq. (\ref{useful_formula}) one obtains that
\begin{eqnarray}
&&\int_\epsilon^0 du \int_0^L dx P_L(u_m, x_m, u,x,u_L=0) \\
&&= \epsilon^2 \int_0^{L-x_m} dx E(u_m,L-x_m,0,x) \int_0^{x_m} dx E(u_m,x_m,0,x) + {\cal O}(\epsilon^3) \;.\nonumber 
\end{eqnarray}
Using the above scaling forms (\ref{scaling_direct}) one obtains that
\begin{eqnarray}
\int_\epsilon^0 du \int_0^L dx P_L(u_m, x_m, u,x,u_L=0) = \epsilon^2 \frac{1}{u_m^6} P_s\left( \frac{u_m}{x_m^2}\right)  P_s\left( \frac{L-u_m}{x_m^2}\right) +  {\cal O}(\epsilon^3) \;,
\end{eqnarray}
where the Laplace transform of $P_s(z)$ is given by
\begin{eqnarray}
\int_0^\infty dz e^{-p z} P_s(z) = \tilde E(p,u=0,q=0) = \frac{\sqrt{p}}{\sinh{(\sqrt{p})}} \;.
\end{eqnarray}
By inverting the Laplace transform, one obtains
\begin{eqnarray}\label{expansion_g}
P_s(z) = 2\pi^2 \sum_{m=0}^\infty (-1)^{m+1}
m^2 e^{-m^2 \pi^2 z}  \;.
\end{eqnarray}
Finally, after normalization, one obtains
\begin{eqnarray}\label{joint_excursion}
P_L^{\rm ex}(u_m, x_m) = \frac{8 \pi^{9/2} L^{3/2}}{u_m^6}
\sum_{m,n=0}^\infty (-1)^{m+n} m^2 n^2 e^{-m^2 \pi^2
  \frac{x_m}{u_m^2}} e^{-n^2 \pi^2
  \frac{L-x_m}{u_m^2}} \;,
\end{eqnarray}  
which gives back the formula obtained by Majumdar {\it et al.} in Ref. \cite{satya_yor} (see their
formula (8)). In particular, by integrating over $u_m$, one obtains the marginal distribution of $x_m$ as
\begin{eqnarray}
&&{\mathcal P}_L^{\rm ex}(x_m) = \frac{1}{L}\tilde {\mathcal P}^{\rm ex}\left(\frac{x_m}{L} \right)  \;, \\
&& \tilde {\mathcal P}^{\rm ex}(x) = 3 \sum_{n,m=0}^\infty (-1)^{m+n} \frac{m^2 n^2}{\left[ n^2 x + m^2(1-x) \right]^{5/2}} \;.
\end{eqnarray}

\subsubsection{Meander}

Here we consider Brownian meanders, which are similar to excursions except that in that case the endpoint $u_L$ is free, see Fig. \ref{fig_constrained} d). 
Thus one would impose the minimum to be $0$ and integrate over $u_L$. Following the same limiting procedure as above one has
\begin{eqnarray}
P_L^{\rm mea}(u_m, x_m) = \lim_{\epsilon \to 0}
\frac{\int_\epsilon^{u_m} du_L \int_\epsilon^0 du \int_0^L dx P_L(u_m, x_m,
  u,x,u_L)}{\int_\epsilon^{u_m} du_L \int_0^\infty du_m \int_0^L dx_m 
  \int_\epsilon^0 du \int_0^L dx P_L(u_m, x_m, u,x,u_L)} \;.
\end{eqnarray} 
Performing the same analysis as before, and using again the identity
in Eq. (\ref{useful_formula}) one obtains that 
\begin{eqnarray}
&&\!\!\!\!\!\!\!\!\!\!\!\!  \int_\epsilon^{u_m} du_L \int_\epsilon^0 du \int_0^L dx P_L(u_m, x_m,
u,x,u_L) \nonumber \\
&&\!\!\!\!\!\!\!\!\!\!\!\!  = \epsilon \int_0^{u_m} du_L \int_0^{L-x_m} dx E(u_m-u_L,L-x_m,0,x)
\int_0^{x_m} dx E(u_m,x_m,0,x) + {\cal O}(\epsilon^3)  \;.
\end{eqnarray}
Therefore, after straightforward algebra, one has here
\begin{eqnarray}
&&\int_\epsilon^{u_m} du_L \int_\epsilon^0 du \int_0^L dx P_L(u_m, x_m,
u,x,u_L) =\epsilon \frac{1}{u_m^4} P_s\left( \frac{u_m}{x_m^2}\right)
P_t\left( \frac{L-u_m}{x_m^2}\right) \;,
\end{eqnarray}
where the Laplace transform of $P_t(z)$ is given by
\begin{eqnarray}
&& \!\!\!\!\!\!\!\!\!\!\!\!  \!\!\!\!\!\!\!\!\!\!\!\!  \int_0^\infty dz e^{-p z} P_t(z) = \sqrt{p} \int_0^1 dv
\sinh{(\sqrt{p}(1-v))} \int_0^v
\frac{du}{(\sinh{(\sqrt{p}(1-v+u))})^2}  = \frac{\tanh{\left(\frac{\sqrt{p}}{2}
    \right)}}{\sqrt{p}} \;.
\end{eqnarray}
Inverting the Laplace transform yields finally 
\begin{eqnarray}
P^{\rm mea}_L(u_m,x_m) = \frac{4 \pi^{5/2} L^{1/2}}{u_m^4} \sum_{m,n=1}^\infty
((-1)^{m+n}-(-1)^n)n^2 e^{-n^2\pi^2\frac{x_m}{u_m^2}}  e^{-m^2 \pi^2
  \frac{L-x_m}{u_m^2}} \;,     
\end{eqnarray}
which gives back the formula obtained by Majumdar {\it et al.} in Ref. \cite{satya_yor} (see their
formula (19)). In particular, by integrating over $u_m$, one obtains the marginal distribution of $x_m$ as
\begin{eqnarray}
&&{\mathcal P}_L^{\rm mea}(x_m) = \tilde {\mathcal P}^{\rm mea}\left(\frac{x_m}{L} \right)  \;, \\
&& \tilde {\mathcal P}^{\rm mea}(x) = \sum_{n,m=1}^\infty ((-1)^{m+n} - (-1)^n)\frac{n^2}{\left[ n^2 x + m^2(1-x) \right]^{3/2}} \\
&& = 2 \sum_{m=0,n=1}^\infty (-1)^{n+1} \frac{n^2}{\left[n^2 x + (2m+1)^2(1-x) \right]^{3/2}} \;,
\end{eqnarray}
where in the last equation we have used that only the terms where $m$ is odd contribute, {\it i.e} $(-1)^{m+n}-(-1)^n = 0$ when $m$ is even.

We have thus shown that the RSRG allows one to recover in a rather
simple way standard results for various constrained Brownian
motions. We now extend this method to other stochastic processes and
consider the case of Bessel Processes. 

\section{Bessel Processes and more}

\subsection{RSRG for more general process}

The RSRG was extended to more general processes in Ref. \cite{pldmonthus}, and we refer to that paper for all details. The generalization of the finite size measure (\ref{fs}) now reads:
\begin{eqnarray} \label{fsg}
&& E^-_\Gamma(0,u_0;u_1,x_1) E^+_\Gamma(u_{1},x_{1},L,u_L) +
\\
&& \fl \sum_{k=1}^\infty E^-_\Gamma(0,u_0;u_1,x_1) \prod_{j=1}^k 
B^+_\Gamma(u_{2 j-1},x_{2j-1};u_{2j},x_{2j}) B^-_\Gamma(u_{2j},x_{2j};u_{2j+1},x_{2j+1}) 
E^+_\Gamma(u_{2 k+1},x_{2k+1},L,u_L) \nonumber \;, 
\end{eqnarray}
which has still a block structure, but with slightly more general blocks. The index ``$-$'' refers to an ascending bond, {\it i.e.} where $u_x$ is on average an increasing function, while ``$+$'' refers to descending bonds. This distinction was not needed and suppressed in (\ref{fs}) for the symmetric BM, but it is useful in general. As compared to Ref. \cite{pldmonthus}, we have reversed the orientation $\pm$. This amounts to the reflection $u(x) \to -u(x)$ w.r.t. the notations there, or, in other words to consider here $u(0^-)=u(L^+)=-\infty$. This implies that the $k=0$ term (the first one) in the large $\Gamma$ limit yields the distribution of the maximum (rather than the minimum there). RSRG equations for these blocks were derived there, which we will not reproduce here, and only a small class of solutions were identified (presumably many more remain to be discovered). 

As an example, the following class of real valued Langevin process
$u_x$ was found to provide solutions of the RSRG equations, and was
studied in Ref. \cite{pldmonthus}:
\begin{eqnarray}
&& du_x = - W'(u_x) + dB_x   \label{langevin} \;,
\end{eqnarray} 
with $\overline{dB_x^2}=2 dx$ a Brownian motion. It represents
processes which undergo diffusion in a given potential $W(u)$ with $u
\in ]-\infty,+\infty[$. The explicit form of the blocks, for any
    $\Gamma$, was obtained in Ref. \cite{pldmonthus}, and we need here
    only the form of the edge blocks, given in Laplace w.r.t. $x_1-x_0$ as: 
\begin{eqnarray} \label{soluE} 
&& E^\pm_{\Gamma}(u_0,u_1,p)=e^{-\frac{1}{2} (W(u_1)-W(u_0))} \tilde E^\pm_{\Gamma}(u_0,u_1,p) \;, \\
&& \tilde E^-_{\Gamma}(u_0,u,p)  = \tilde E^+_{\Gamma}(u,u_0,p)  =   
\exp{\left(\int_{u}^{u_0} dv \partial_1 \ln K(v,v-\Gamma,p) \right)}
\;, \nonumber
\end{eqnarray}
where
\begin{eqnarray}
&&  K(u,v,p) =\frac{1}{w(p)} (\Phi_1(u,p) \Phi_2(v,p) - \Phi_2(u,p) \Phi_1(v,p)) \;,
\end{eqnarray}
where $\Phi_i(u,p)$ are the two independent solutions of the associated Schr\"odinger problem:
\begin{eqnarray} \label{eqphi} 
&& \partial^2_u - (p + V(u))]\Phi(u,p)=0 \quad , \quad V(u)=\frac{1}{4} W'(u)^2 -\frac{1}{2} W''(u) \;,
\end{eqnarray}
and $w(p)$ their Wronskian, $w(p) = \Phi_1(u,p) \partial_u \Phi_2(u,p)
- \Phi_2(u,p) \partial_u \Phi_1(u,p)$. Bulk bonds have a similar expression in terms of the kernel $K(u,v,p)$. Note that in the limit $\Gamma \to \infty$ the quantity:
\begin{eqnarray} \label{dlog} 
&&  \partial_1 \ln K(v,v-\Gamma,p) = \frac{\Phi'_1(v,p) \Phi_2(v-\Gamma,p) - \Phi'_2(v,p) \Phi_1(v-\Gamma,p)}{\Phi_1(v,p) \Phi_2(v-\Gamma,p) - \Phi_2(v,p) \Phi_1(v-\Gamma,p)} \;,
\end{eqnarray}
automatically becomes equal to $\Phi'(v,p)/\Phi(v,p)$ where $\Phi(v,p)$ is the linear combination of $\Phi_1$ and $\Phi_2$ orthogonal to the one which blows up at $v \to -\infty$. This is what is expected for a process $u_x \in ]-\infty,+\infty[$.

\subsection{Extreme value statistics of the Bessel Process} 

\begin{figure}[h]
\centering
\includegraphics[width=0.8\linewidth]{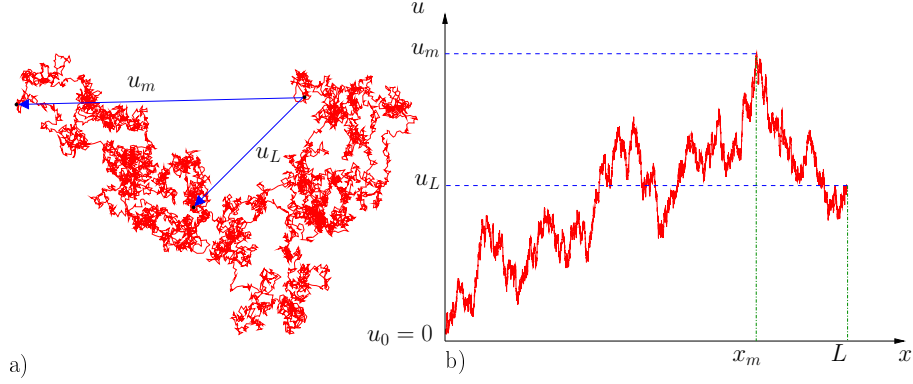}
\caption{Bessel process in dimension $d=2$, {\it i.e.} $\nu = 0$. a) :
  Brownian motion in dimension $d=2$. b) The Bessel process with index
  $\nu=0$ corresponds to the radius of the $2$-dimensional Brownian motion.}\label{fig_bessel}
\end{figure}

We now call $u_x$ the Bessel process, {\it i.e.} the radius of the
$d$-dimensional Brownian motion $u_x=\sqrt{\sum_{i=1} (B_x^i)^2}$
where the $B_x^i$ are $d$ independent BM with $\overline{(dB^i_x)^2}=2
dx$, $i=1, \dots, d$. As is well known it satisfies the one dimensional Langevin equation $du_x = dB_x + (d-1)/u_x$, $u_x>0$, with $\overline{dB_x^2}=2 dx$. Hence it is of the type (\ref{langevin}) with a potential $W(u)=-r \ln |u|$ with $r=d-1$. This allows a generalization for real values of $d$. 

Consider now a Bessel process (BP) starting at $u=u_0$ at $x=0$, and ending at $u_L$ at $x=L$. Let us call 
$P_L(u_m,x_m,u_L|u_0)$ the joint probability distribution that the BP $u_x$ starting at $u_0$ has minimum $u_m$ at position $x_m$ {\it and} ends up at $u_L$. It is given by the last RSRG block in (\ref{fsg}):
\begin{eqnarray}
P_L(u_m,x_m,u_L|u_0) = \lim_{\Gamma \to \infty} E^-_{\Gamma}(u_0,u_m,x_m) E^+_{\Gamma}(u_m,u_L,L-x_m)  \;.
\end{eqnarray}
It satisfies the normalization condition:
\begin{eqnarray} \label{norm2} 
\int_0^L dx_m \int^{u_0}_{-\infty} du_m \int_{u_m}^0 du_L P_L(u_m,x_m,u_L|u_0) = 1 \;,
\end{eqnarray}
which we will check below explicitly on our solution. In the case of
the BP, there is a subtlety with applying the recipe given in Ref. \cite{pldmonthus}
and the solution (\ref{soluE}). Indeed the process remains on the interval $u_x \in [0,+\infty[$. 
Hence $u=0$ has a special role, related to the returns to the origin. 

Let us first examine the naive solutions of (\ref{eqphi}) for the potential $V(u)=\frac{1}{4} W'(u)^2 -\frac{1}{2} W''(u)=(d-1)(d-3)/(4 u^2)$ for the BP. The two independent solutions read:
\begin{eqnarray}
\Phi_1(u,p) = \sqrt{u} K_{1-\frac{d}{2}}(u \sqrt{p}) \quad , \quad \Phi_2(u,p) = \sqrt{u} I_{1-\frac{d}{2}}(u \sqrt{p}) \;,
\end{eqnarray}
where $I_\rho(x)$ and $K_\rho(x)$ are modified Bessel functions of the
first and second kind respectively. Note that for $d=1$ one recovers
the result for the symmetric BM, in which case $W=0$ and  
$\Phi_1(u,p) \sim e^{- u \sqrt{p}}$, $\Phi_2(u,p) \sim \sinh(u \sqrt{p})$, up to unimportant $p$-dependent normalization which is cancelled by the Wronskian. For $d<2$ the above solutions behave as $u \to 0^-$ and for fixed $p>0$:
\begin{eqnarray}
\!\!\!\!\!\!\!\!\!\!\!\!  \Phi_1(u,p) = 2^{-d/2} p^{\frac{d-2}{4}} u^{\frac{d-1}{2}} \Gamma(1-d/2) \quad , \quad 
\Phi_2(u,p) = 2^{-1+d/2} p^{\frac{2-d}{4}} u^{\frac{3-d}{2}} /\Gamma(2-d/2) \;,
\end{eqnarray}
which, upon a rescaling by $p^{(d-1)/2}$ gives also the $p=0$ limit. In the
above expression, $\Gamma(x)$ is the Gamma function. In that limit
$p=0$ these two solutions 
correspond to the general expression \cite{pldmonthus}:
\begin{eqnarray}
&&  \Phi_1(u,p=0) = e^{- \frac{1}{2} W(u)} \quad , \quad \Phi_2(u,p=0) = e^{- \frac{1}{2} W(u)} \int_0^u du' e^{W(u')} \;.
\end{eqnarray}

The process being only defined for $u_x \geq 0$ the question arises of
the correct solution which is selected in the limit $\Gamma \to
+\infty$. First of all the above formulae in Eq. (\ref{soluE}) must be
replaced by: 
\begin{eqnarray}
&& E^\pm_{\Gamma}(u_0,u_1,p)=e^{-\frac{1}{2} (W(u_1)-W(u_0))} \tilde E^\pm_{\Gamma}(u_0,u_1,p) \;, \\
&& \tilde E^-_{\Gamma}(u_0,u,p)  = \tilde E^+_{\Gamma}(u,u_0,p)  =   \exp{\left(\int_{u}^{u_0} dv \partial_1 \ln K(v,\max(0,v-\Gamma),p) \right)} \;,
\end{eqnarray}
and the question of the value for $K(u,0,p)$ arises. One way to regularize the problem would be to add
near $u_0>0$ a very steep barrier (e.g. $W(u)=e^{u_0/|u|}$) and consider $K(u,v,p)$ in the limit $|v| \gg u_0$ both going to zero.
A short-cut to treat that barrier problem is to notice that this is equivalent to forget the barrier and ask for reflecting boundary conditions at $v=0$ for the process $u_x$. We must also ask that as $u \to 0$ the current vanish so that:
\begin{eqnarray}
J_S = \left( \partial_u  - \frac{1}{2} W'(u) \right) K(u,0,p) = \left( \partial_u  - \frac{d-1}{2 u} \right) K(u,0,p) = 0 \;.
\end{eqnarray} 
It is then easy to find that the proper solution is $K(u,0,p) \sim
\cosh(u \sqrt{p})$ for the symmetric BM ({\it i.e.} when $W(u)=0$ {\it
  i.e.} the $d=1$ BP becomes the reflected BM see below), while for
the BP we find: 
\begin{eqnarray}
K(u,0,p) = A \left( \frac{2}{\pi} \sin{\left(d \frac{\pi}{2}\right)} \Phi_1(u) +  \Phi_2(u) \right) = \sqrt{u} I_{-1+\frac{d}{2}}(u \sqrt{p}) \;.
\end{eqnarray}
where $A$ is some constant. We may understand this from a limit case of the regularization by a barrier at scale $u_0$ writing:
\begin{eqnarray}
K(u,v,p) \sim \Phi^{u_0}_1(u) \Phi^{u_0}_2(v) - \Phi^{u_0}_2(u) \Phi^{u_0}_1(v) \;,
\end{eqnarray} 
where $\Phi^{u_0}_{1,2}(u)$ are respectively decaying and exploding at $u=+\infty$, 
and showing that irrespective of the form of the barrier it reflects hence:
\begin{eqnarray}
\lim_{|v| \gg u_0 \to 0} \frac{ \Phi^{u_0}_2(v) }{\Phi^{u_0}_1(v)}  =
-  \frac{2}{\pi} \sin{\left (d \frac{\pi}{2}\right)} \;.
\end{eqnarray} 
 
Using this result we now find that the large $\Gamma$ limit is given by:
\begin{eqnarray}
&& \lim_{\Gamma \to \infty} E^-_{\Gamma}(u_0,u_m,p) = e^{-\frac{1}{2} (W(u_m)-W(u_0))}
\frac{\frac{2}{\pi} \sin(\frac{\pi d}{2}) \Phi_1(u_0,p) + \Phi_2(u_0,p)}{\frac{2}{\pi} \sin(d \frac{\pi}{2})  \Phi_1(u_m,p) 
+  \Phi_2(u_m,p)} \\
&& = \left(\frac{u_m}{u_0}\right)^{\frac{d-2}{2}} \frac{ \frac{2}{\pi} \sin( \frac{\pi d}{2}) K_{1-\frac{d}{2}}(u_0 \sqrt{p}) +   I_{1-\frac{d}{2}}(u_0 \sqrt{p}) }{\frac{2}{\pi} \sin( \frac{\pi d}{2}) K_{1-\frac{d}{2}}(u_m \sqrt{p}) +    I_{1-\frac{d}{2}}(u_m \sqrt{p})} \\
&& = \left(\frac{u_m}{u_0}\right)^{\frac{d-2}{2}} \frac{I_{-1+\frac{d}{2}}(u_0 \sqrt{p})}{I_{-1+\frac{d}{2}}(u_m \sqrt{p})} \;,
\end{eqnarray}
which is found to hold for any $d$. One finds similarly:
\begin{eqnarray}
&& \lim_{\Gamma \to \infty} E^+_{\Gamma}(u_m,u_L,p) 
= \left(\frac{u_L}{u_m}\right)^{\frac{d}{2}} \frac{I_{-1+\frac{d}{2}}(u_L \sqrt{p})}{I_{-1+\frac{d}{2}}(u_m \sqrt{p})} \;.
\end{eqnarray}
One can check the normalization (\ref{norm2}), {\it i.e.} in Laplace, {\it i.e.} one writes the product:
\begin{eqnarray}
&& \lim_{\Gamma \to \infty} E^-_{\Gamma}(u_0,u_m,p) E^+_{\Gamma}(u_m,u_L,p) 
= \frac{1}{u_m} \frac{u_L^{d/2}}{u_0^{(d-2)/2}} \frac{I_{-1+\frac{d}{2}}(u_0 \sqrt{p}) I_{-1+\frac{d}{2}}(u_L \sqrt{p}) }{I_{-1+\frac{d}{2}}(u_m \sqrt{p})^2} \;.
\end{eqnarray}
Integration over $u_L$ and $u_m$ as described in (\ref{norm2}) yields $1/p$, as required.

For $d=1$, which is just the reflected Brownian (see Fig. \ref{fig_constrained} b) one finds:
\begin{eqnarray}
\!\!\!\!\!\!\!\!\!\!\!\!  \lim_{\Gamma \to \infty} E^-_{\Gamma}(u_0,u_m,p) = \frac{\cosh(u_0 \sqrt{p})}{\cosh(u_m \sqrt{p})} \quad , \quad
\lim_{\Gamma \to \infty} E^+_{\Gamma}(u_m,u_L,p) = \frac{\cosh(u_L \sqrt{p})}{\cosh(u_m \sqrt{p})} \;.
\end{eqnarray}
Similarly for $d=3$, one finds:
\begin{eqnarray}
\!\!\!\!\!\!\!\!\!\!\!\!  \lim_{\Gamma \to \infty} E^-_{\Gamma}(u_0,u_m,p) = \frac{u_m \sinh(u_0 \sqrt{p})}{u_0 \sinh(u_m \sqrt{p})}  \quad , \quad
\lim_{\Gamma \to \infty} E^+_{\Gamma}(u_m,u_L,p) = \frac{u_L \sinh(u_L \sqrt{p})}{u_m \sinh(u_m \sqrt{p})} \;.
\end{eqnarray}
The Inverse Laplace transform now gives, for any $d$:
\begin{eqnarray}
&& \lim_{\Gamma \to \infty} E^-_{\Gamma}(u_0,u_m,x) = \left(\frac{u_m}{u_0}\right)^{\frac{d-2}{2}}
\sum_{n} \frac{2}{u_m^2} i j_{\nu,n} \frac{I_{\nu}(i j_{\nu,n} \frac{u_0}{u_m})}{I'_{\nu}(i j_{\nu,n})} e^{- \frac{j_{\nu,n}^2 x}{u_m^2}}  \;,
\end{eqnarray}
with $I_\nu(z) = e^{-\frac{1}{2} i \pi \nu} J_\nu(i z)$, $\nu=-1 + d/2$, and $0 < j_{\nu,1} < j_{\nu,2}<...$ is the sequence of positive zeroes of $J_\nu$, 
{\it i.e.} $J_\nu(j_{\nu,n})=0$, where $J_\nu(x)$ is a Bessel function
of the first kind. It can be further simplified using standard relations between Bessel functions
\begin{eqnarray}\label{final_egm}
&& \lim_{\Gamma \to \infty} E^-_{\Gamma}(u_0,u_m,x) = \left(\frac{u_m}{u_0}\right)^{\frac{d-2}{2}}
\sum_{n} \frac{2}{u_m^2} j_{\nu,n} \frac{J_{\nu}(j_{\nu,n} \frac{u_0}{u_m})}{J_{\nu+1}(j_{\nu,n})} e^{- \frac{j_{\nu,n}^2 x}{u_m^2}} \;,
\end{eqnarray}
and, similarly:
\begin{eqnarray}\label{final_egp}
&& \lim_{\Gamma \to \infty} E^+_{\Gamma}(u_m,u_L,x) = \left(\frac{u_L}{u_m}\right)^{\frac{d}{2}}
\sum_{n} \frac{2}{u_m^2} j_{\nu,n} \frac{J_{\nu}(j_{\nu,n} \frac{u_L}{u_m})}{J_{\nu+1}(j_{\nu,n})} e^{- \frac{j_{\nu,n}^2 x}{u_m^2}} \;.
\end{eqnarray}
Combining these two expressions (\ref{final_egm}, \ref{final_egp}) yields our final result for the joint distribution of the maximum, its position and the endpoint of the Bessel Process:
\begin{eqnarray}
&& \!\!\!\!\!\!\!\!\!\!\!\! \!\!\!\!\!\!\!\!\!\!\!\!  P_L(u_m,x_m,u_L|u_0)  = \frac{u_L^{d/2}}{u_0^{(d-2)/2}} \frac{4}{u_m^5} \sum_{n,m} j_{\nu,n} j_{\nu,m} \frac{J_{\nu}(j_{\nu,n} \frac{u_0}{u_m})     J_{\nu}(j_{\nu,m} \frac{u_L}{u_m})   }{J_{\nu+1}(j_{\nu,n})   J_{\nu+1}(j_{\nu,m})} e^{- \frac{j_{\nu,n}^2 x_m}{u_m^2}}  e^{- \frac{j_{\nu,m}^2 (L-x_m)}{u_m^2}} \nonumber \;,\\
\end{eqnarray}
with $\nu=-1 + d/2$. 

We now focus on Bessel bridges for which $u_0 = u_L = 0$. Here again one uses a limiting procedure as before and writes
\begin{eqnarray}
P^{\rm Bessel \, bridge}_L(u_m,x_m) = \lim_{\epsilon\to 0} \frac{ P_L(u_m,x_m,u_L=\epsilon|u_0 = \epsilon)}{\int_{\epsilon}^\infty du_m \int_0^L dx_m  P_L(u_m,x_m,u_L=\epsilon|u_0 = \epsilon)  } \;.
\end{eqnarray}
Using the small $x$ behavior $J_\nu(x) \sim x^\nu/(2^\nu \Gamma(\nu+1))$ one obtains
\begin{eqnarray}\label{joint_bessel}
\!\!\!\!\!\!\!\!\!\!\!\! \!\!\!\!\!\!\!\!\!\!\!\!  P^{\rm Bessel \, bridge}_L(u_m,x_m) = \frac{8 L^{\nu+1}}{\Gamma(1+\nu) u_m^{5+2\nu}} \sum_{n,m} \frac{   j_{\nu,n}^{\nu+1} j_{\nu,m}^{\nu+1}    }{J_{\nu+1}(j_{\nu,n})   J_{\nu+1}(j_{\nu,m})} e^{- \frac{j_{\nu,n}^2 x_m}{u_m^2}}  e^{- \frac{j_{\nu,m}^2 (L-x_m)}{u_m^2}} \;.
\end{eqnarray}
For $d=1$, one has $\nu = -1/2$ and $j_{-1/2,n} = (n+\frac{1}{2}) \pi$, one checks that this formula gives back the result for the reflected Brownian motion in Eq. (\ref{joint_reflected}). For $d=3$ one has $\nu = 1/2$ and $j_{-1/2,n} = n \pi$, one checks that this formula gives back the result for the excursion in Eq. (\ref{joint_excursion}). 

On the other hand, if one integrates over $x_m$ one recovers the distribution of $u_m$ as obtained by Gikhman \cite{gikhman} and Kiefer \cite{kiefer} for integer values of the dimension $d$ and later generalized by Pitman and Yor in Ref. \cite{pitman_yor} for non integer $d$ (see \ref{appendix_pityor}). It reads
\begin{eqnarray}\label{marginal_max_bessel}
 \tilde {\rm P}^{\rm Bessel \, bridge}(x) = \frac{d}{dx} \Bigg( \frac{4}{\Gamma(\nu+1) x^{2\nu+2}} \sum_{n=1}^{\infty} \frac{j_{\nu,n}^{2\nu}}{J^2_{\nu+1}(j_{\nu,n})} \exp{\left(-\frac{j_{\nu,n}^2}{x^2} \right)}  \Bigg) \;.
\end{eqnarray}
Now if one integrates instead over $u_m$ one obtains the marginal distribution of $x_m$ as
\begin{eqnarray}\label{marginal_pos_bessel}
&&{\cal P}^{\rm Bessel \, bridge}_L(x_m) = \frac{1}{L} \tilde {\cal P}^{\rm Bessel \, bridge}\left(\frac{x_m}{L}\right) \\
&& \tilde {\cal P}^{\rm Bessel \, bridge}(x) = 4 (1+\nu) \sum_{n,m} \frac{   j_{\nu,n}^{\nu+1} j_{\nu,m}^{\nu+1}    }{J_{\nu+1}(j_{\nu,n})   J_{\nu+1}(j_{\nu,m})   \left[  j_{\nu,n}^2 x + j_{\nu,m}^2 (1-x)     \right]^{\nu+2}} \;.
\end{eqnarray}

Many more quantities could be computed and it remains also to extend these considerations to
finite $\Gamma$ and bulk bonds. We leave this to future studies. 

\section{Continuous time random walks: a new RSRG fixed point}

We now come to the study of the so called continuous time random walks
(CTRW), which was first introduced by Montroll
and Weiss in Ref. \cite{montroll_weiss}. Within this model of
diffusion, the walker performs a usual random walk but has to wait for a certain ''trapping time'' $\tau$ before each jump. The trapping times between each jump
are independent and identically distributed random variables with a common density function $\psi(\tau)$ which has
a power law tail, $\psi(\tau) \propto \tau^{-1-\alpha}$ with $0 <
\alpha < 1$, while the jumps themselves are 
distributed according to a narrow distribution. This type of model was
suggested by 
Scher and Montroll \cite{scher_montroll} to model non-Gaussian
transport of electrons in disordered systems and since 
it has been widely used to describe phenomenologically anomalous
dynamics in various complex systems \cite{bouchaud_georges,
  klafter_review}. Indeed, for 
$\alpha < 1$, the mean trapping time between two successive jumps is
infinite and hence CTRW is characterized by a subdiffusive behavior,
with a dynamical exponent $z 
= 2/\alpha > 1$, and non-Gaussian statistics. While CTRW has been widely
studied \cite{bouchaud_georges, klafter_review}, 
it seems that the extreme statistics of these processes 
have not been studied in detail. It is the purpose of this section to
study them using RSRG. 

We now trade the variable $\tau$ for $x$. The CTRW described above
generates a landscape $u(x)$. Performing the first stages of the RSRG
method as described in Ref. \cite{pldmonthus}, one finds that the
bonds in the renormalized landscape at scale $\Gamma$ acquire a broad
distribution, with a typical length $\Gamma^{2/\alpha}$, while the
jumps in $u$ remain of order $\Gamma$. One can now search for a fixed
point of the RSRG which has precisely this scaling with $\Gamma$. This
is done in \ref{appendix_ctrw} and we find that these new fixed points
can be simply obtained (up to some subtelties explained below) from the Brownian case by substituting $p$ to $p^\alpha$ ($\alpha =1$
corresponding to Brownian motion) in the RSRG-blocks of the BM.

Therefore it is natural to consider the joint probability density $Q_L(u_m,x_m,u_L)$ that the free
CTRW just {\it arrives} at $u_L$ at time $L$ and has maximum value
$u_m$ at position $x_m$. From similar considerations as in
Eq. (\ref{last_block}), and the discussion in \ref{appendix_ctrw}, it is given by 
\begin{eqnarray}\label{last_block_ctrw}
&& Q_L(u_{m},x_{m},u_L) = \lim_{\Gamma \to \infty} 
\Gamma^{2/\alpha} E_\Gamma(F_1,x_m) E_\Gamma(F_2,L-x_m) \;,
\end{eqnarray}
with $F_1=u_{m}$ and
$F_2=u_{m}-u_L$. Using the result (\ref{eg}) for the
edge blocks together with the substitution $p$ to $p^\alpha$ (see also Eq. \ref{sol_edge_bond_app} in \ref{appendix_ctrw}) one obtains in the large $\Gamma$ limit:
\begin{eqnarray}
\lim_{\Gamma \to \infty} \Gamma^{1/\alpha} E_\Gamma(F,p) = e^{-F p^{\alpha/2}} \;,
\end{eqnarray}
which gives after inverse Laplace transformation :
\begin{eqnarray}\label{edge_ctrw}
\lim_{\Gamma \to \infty} \Gamma^{1/\alpha} E_\Gamma(F,l) = F^{-2/\alpha} {\cal L}_{\alpha/2} (l F^{-2/\alpha}) \;,
\end{eqnarray}
where ${\cal L}_\nu(x)$ is a one-sided L\'evy stable pdf of index $\nu = \alpha/2$, whose
Laplace transform is $\int_0^\infty e^{-p x} {\cal L}_\nu(x) =
e^{-p^\nu}$. The function ${\cal L}_\nu(x)$ can be represented in
terms of Fox $H$ functions \cite{schneider, mainardi, barkai_fox} or,
alternatively, can be 
written as a power series~\cite{pollard} 
\begin{eqnarray}\label{levy_series}
{\cal L}_\nu(x) = \frac{1}{\pi} \sum_{n=1}^\infty (-1)^{n+1} x^{-\nu
  n-1} \frac{\Gamma(n \nu +1)}{n!} \sin{(n \pi \nu)} \;, \; x > 0  \;,
\end{eqnarray}
from which one obtains the asymptotic behavior for large argument $x
\gg 1$ as ${\cal L}_\nu(x) \sim \nu \Gamma(1-\nu)^{-1} x^{-(1+\nu)}$,
where we have used $\Gamma(\nu)
\Gamma(1-\nu) = \pi/\sin{(\nu \pi)}$. For small argument, it behaves
as~\cite{schneider} 
\begin{eqnarray}
{\cal L}_\nu(x) \sim x^{-\sigma} e^{-\kappa x^{-\tau}} \;,
\end{eqnarray}
where $\tau = \nu/(1-\nu)$, $\kappa = (1-\nu) \nu^{\nu/(1-\nu)}$,
$\sigma = (2-\nu)/(2(1-\nu))$. 

For $\nu = 1/2$, one has 
\begin{eqnarray}
{\cal L}_{1/2}(x) = \frac{1}{2 \sqrt{\pi} x^{3/2}} e^{-1/(4x)} \;,
\end{eqnarray}
and thus the formula (\ref{edge_ctrw}) yields back the result for
Brownian motion in Eq. (\ref{edge_brownian}). For other values of
$\nu$, a representation of ${\cal L}_\nu(x)$ in terms of other simpler
special functions is usually obtained by examining the series expansion in
Eq. (\ref{levy_series}) and using properties of gamma-~functions. For any rational value of $\nu$, it was shown in
Ref. \cite{scher_montroll} that ${\cal L}_\nu(x)$ can be expressed in
terms of a finite sum of hypergeometric functions (for instance,
${\cal L}_{1/4}(x)$ is given in Ref. \cite{barkai_fox}). Here we
simply quote another interesting simple case corresponding to $\nu =
1/3$, {\it i.e.} $\alpha = 2/3$ for which one has 
\begin{eqnarray}
{\cal L}_{1/3}(x) = \frac{1}{(3x^4)^{1/3}} {\rm
  Ai}\left[\left(3x\right)^{-1/3}\right] \;,
\end{eqnarray}  
where ${\rm Ai}(x)$ is the Airy function.

Now the probability that the free CTRW {\it ends} in $u_L$ at time $L$
and has maximum value $u_m$ at position $x_m$ is obtained from
$Q_L(u_{m},x_{m},u_L)$ as a convolution : 
\begin{eqnarray}
 P_L(u_{m},x_{m},u_L) = \int_{x_m}^L d L' Q_{L'}(u_{m},x_{m},u_L) \Psi(L-L') \;,
\end{eqnarray}
where $\Psi(l) = \int_l^\infty \psi(x') dx'$ is the probability of no jump in the time interval $[0,l]$. Its Laplace transform $\Psi(p)$ behaves, for small $p$, as
$\Psi(p) = A  p^{\alpha-1}$ with $A=1$, see \ref{appendix_ctrw}. And therefore one obtains the double Laplace transform of $P_L(u_{m},x_{m},u_L)$ with respect to $x_m$ and $L$ as
\begin{eqnarray}\label{start_expr_ctrw}
 \int_0^{L} dx_m e^{-q x_m} \int_0^\infty dL e^{-p L}  P_L(u_{m},x_{m},u_L) = e^{-(p+q)^\nu u_m} e^{-p^\nu (u_m-u_L)} p^{2\nu -1} \;,
\end{eqnarray}
 where $\nu = \alpha/2$. From Eq. (\ref{start_expr_ctrw}), one checks
 that one recovers the correct expression for the propagator
 $W(u_L,L)$, namely the probability that the free CTRW {\it ends} in
 $u_L$ at time $L$. It is given by
\begin{eqnarray}
W(u_L,L) = \int_{\max(u_L,0)}^\infty d u_m \int_{0}^L dx_m P_L(u_{m},x_{m},u_L) \;.
\end{eqnarray}
One obtains its Laplace transform with respect to $L$ as
\begin{eqnarray}
W(u_L,p) = \int_0^\infty dL e^{-pL} W(u_L,L) = \frac{1}{2} p^{\nu-1}
e^{-p^\nu |u_L| }  \;,
\end{eqnarray}
which yields the correct expression for the Laplace transform (with
respect to time) of the propagator \cite{klafter_review}. One can
invert the Laplace transform to obtain 
\begin{eqnarray}\label{propag_ctrw}
W(u_L,L) = \frac{L}{2\nu |u_L|^{1+1/\nu}} {\cal
  L}_\nu\left(\frac{L}{|u_L|^{1/\nu}} \right) \;, \; \nu = \alpha/2 \;.
\end{eqnarray}  
For $\alpha = 1$, it yields back the expression in Eq. (\ref{propag_bm}) and
for $\alpha = 2/3$ Eq. (\ref{propag_ctrw}) takes the simple form
\begin{eqnarray}
W(u_L,L) = \frac{3^{2/3}}{2 L^{1/3}} {\rm
  Ai}\left(\frac{|u_L|}{(3 L)^{1/3}} \right)  \;,
\end{eqnarray}
with ${\rm Ai(x)} \sim (3^{2/3} \Gamma(2/3))^{-1} - x (3^{1/3}
\Gamma(1/3))^{-1}$ for $x \to 0$ and ${\rm Ai(x)} \sim (2 x^{1/2}
\pi^{1/2})^{-1} e^{-2x^{2/3}/3}$ for $x \to \infty$.

We consider the case of the free CTRW where one integrates over the final position $u_L$ between $-\infty$ and $u_m$. In that case one immediately gets, from Eq. (\ref{start_expr_ctrw}) the expression for the joint distribution $P^{\rm free}_L(u_m,x_m)$ as
\begin{eqnarray}\label{joint_free_ctrw}
 P^{\rm free}_L(u_m,x_m) = \theta(u_m) \theta(L-x_m) \frac{1}{\Gamma[1-\nu]} \frac{1}{u_m^{1/\nu} (L-x_m)^\nu} {\cal L}_\nu \left(\frac{x_m}{u_m^{1/\nu}} \right) \;.
\end{eqnarray}
In particular, from Eq. (\ref{joint_free_ctrw}) one obtains the marginal distribution of $x_m$ as
\begin{eqnarray}
 {\cal P}_L^{\rm free}(x_m) = \frac{1}{L} \tilde {\cal P}^{\rm
 free}\left(\frac{x_m}{L}\right) \;, \;  \tilde {\cal P}^{\rm
 free}(x) = \frac{\sin{\nu \pi}}{\pi} \frac{1}{x_m^{1-\nu} (1-x_m)^\nu} \;,
\end{eqnarray}
where we have used $\int_0^\infty dy y^{-\nu} {\cal L}_\nu(y) =
1/\Gamma[1+\nu]$.
Similarly, one has the marginal distribution of the maximum $u_m$ as 
\begin{eqnarray}
\Prm^{\rm free}_L(u_{m})  = \frac{1}{L^{\nu}} \tilde \Prm^{\rm
    free}\left(\frac{u_m}{L^{\nu}} \right) \;, \; \tilde \Prm^{\rm
    free} (x) = \theta(x)  \frac{1}{\nu} x^{-(1+1/\nu)}  {\cal L}_\nu
    \left(x^{-1/\nu} \right).\;
\end{eqnarray} 
Of course, following the same lines as done previously for Brownian
motion, one could also consider various constrained CTRW: this is left
for future investigations. 

\section{Conclusion}

In conclusion, we have shown that the RSRG is a quite powerful method
to compute the extreme statistics of various physically relevant
stochastic processes. By exploiting the solution of the RSRG equations
found in Ref. \cite{sinaiRSRG, pldmonthus}, we have studied the extreme
statistics of the one-dimensional Brownian motion (BM) as well as the
Bessel process (BP), {\it i.e.} the radius of the $d$-dimensional Brownian
motion. For the BM, we have shown that it allows to recover,
in a rather different way, standard results for the Brownian motion
and its variants, including Brownian bridge, excursion, meander as
well as for the reflected Brownian motion. For the BP, we have
recovered in a simpler way the results of Pitman and Yor for the
distribution of the 
maximum, and obtained also the distribution of its position. 

We have then extended this  method to study the extreme statistics of the
Continuous Time Random Walk (CTRW), which we have shown to correspond
to a new fixed point of the RSRG transformation. Although we have
restricted our analysis to the extreme statistics of the free CTRW,
various cases of constrained CTRW can be straightforwardly studied
following the analysis presented above. Similarly, the study of the
dynamics in a disordered energy landscape generated by a CTRW could be
done in principle from the analysis of the new fixed point of the RSRG
exhibited here.

\newpage

\appendix

\section{RSRG equations and fixed point for bulk and edge bonds}\label{appendix_edge}

Here we recall the RSRG equation for the probability distributions of barriers and lengths in a symmetric landscape denoted $P_\Gamma(F,l)$ for bulk bonds and $E_\Gamma(F,\ell)$ for edge bonds. When convenient we use the notation $\zeta=F-\Gamma$. For bulk bonds it reads:
\begin{eqnarray} \label{rsrgp}
(\partial_\Gamma - \partial_\zeta) P_\Gamma(\zeta,\ell) = \int_{\ell_1+\ell_2+\ell_3=\ell} P_\Gamma(0,\ell_2)  \int_0^\zeta d\zeta' P_\Gamma(\zeta',\ell_1) P_\Gamma(\zeta-\zeta',\ell_3) \;,
\end{eqnarray}
and for edge bonds one has:
\begin{eqnarray}
&& \partial_\Gamma E_\Gamma(F,\ell) = - P_\Gamma(0) E_\Gamma(F,\ell) \nonumber \\
&& + \int_{\ell_1+\ell_2+\ell_3=\ell} \int_0^\infty dF_1 \int_0^\infty d\zeta_3
E_\Gamma(F_1,\ell_1) P_\Gamma(0,\ell_2) P_\Gamma(\zeta_3,\ell_3)
\delta(F-(F_1+\zeta_3))  \;.
\end{eqnarray}
In terms of Laplace variable $p$ with respect to $l$, one obtains:
\begin{eqnarray}\label{eq_edge_bond1}
&& \partial_\Gamma E_\Gamma(F,p) = - P_\Gamma(0) E_\Gamma(F,p) +
  \int_0^F dF_1  
E_\Gamma(F_1,p) P_\Gamma(0,p) P_\Gamma(F-F_1,p) \;,
\end{eqnarray}
and we recall:
\begin{eqnarray}
&& P_\Gamma(\zeta,p) = a_\Gamma(p) e^{- \zeta b_\Gamma(p)} \\ \label{ab}
&& a(p) = \frac{\sqrt{p}}{\sinh(\Gamma \sqrt{p})}  \quad , \quad b_\Gamma(p)=\sqrt{p} \coth(\Gamma \sqrt{p}) \;.
\end{eqnarray}
It is straightforward to check that:
\begin{eqnarray}\label{sol_edge_bond_app}
&& E_\Gamma(F,p) = \Gamma^{-1} e^{- F \sqrt{p} \coth(\Gamma \sqrt{p})}
\end{eqnarray}
is a solution of Eq. (\ref{eq_edge_bond1}).

\section{A new fixed point for the RSRG}\label{appendix_ctrw}

We now seek a solution of (\ref{rsrgp}) with the scaling form:
\begin{eqnarray}
P_\Gamma(\zeta,\ell) = \Gamma^{-1-\frac{2}{\alpha}} Q_\Gamma\left(\eta=\frac{\zeta}{\Gamma},\lambda = \frac{\ell}{\Gamma^{2/\alpha}}\right) \;,
\end{eqnarray}
which, as discussed in the text corresponds to a CTRW energy landscape with index $\alpha$ (we remind that one recovers the standard BM for $\alpha=1$). Consider $Q_\Gamma(\eta,\tilde p)$ the Laplace transform of $Q_\Gamma(\eta,\lambda)$ w.r.t. $\lambda$ only. It 
satisfies the flow and fixed point equation:
\begin{eqnarray}
0 \equiv \Gamma \partial_\Gamma Q =  Q + (1+\eta) \partial_\eta \hat Q - \frac{2}{\alpha} p \partial_p \hat Q+
Q(0,p) Q (.,p) *_{\eta} Q(.,p) \;.
\end{eqnarray}
One can check that the solution of the fixed point equations take the form:
\begin{eqnarray}
Q(\eta,\tilde p) = a_{\Gamma=1}(\tilde p^\alpha) e^{- \eta b_{\Gamma=1}(\tilde p^\alpha)}
\end{eqnarray}
where the functions $a_\Gamma(p)$ and $b_\Gamma(p)$ are given in (\ref{ab}). Reexpressed
in the Laplace variable $p$ associated with $\ell$ we thus find:
\begin{eqnarray}\label{bulk_sol_ctrw}
&& P_\Gamma(\zeta,p) = \frac{p^{\alpha/2}}{\sinh(\Gamma p^{\alpha/2})}
e^{- \zeta p^{\alpha/2} \coth(\Gamma p^{\alpha/2}) } \;.
\end{eqnarray}
This is a new class of RSRG fixed points which corresponds to broad distributions of bond length:
\begin{eqnarray}
&& P_\Gamma(\ell) = LT^{-1}_{p \to \ell} \frac{1}{\cosh(\Gamma p^{\alpha/2})} \;.
\end{eqnarray}
While there is a typical bond length, $\ell_{\rm typ} \sim \Gamma^{2/\alpha}$ the average bond length is
infinite as the distribution does not have a first moment. Expanding at small $p$ one has $P_\Gamma(p)=1-\frac{1}{2} \Gamma^2 p^\alpha + ..$ from which we find the tail $P_\Gamma(\ell) \sim \Gamma^2/\ell^{1+\alpha}$ at large $\ell$. 

For the edge bonds one finds similarly that
\begin{eqnarray} \label{sol_edge_bond_app}
&& E_\Gamma(F,p) = \Gamma^{-1/\alpha} e^{- F p^{\alpha/2} \coth(\Gamma p^{\alpha/2})}
\end{eqnarray}
is a solution (in Laplace) of Eq. (\ref{eq_edge_bond1}). Hence we see, that, as claimed in the
text the CTRW fixed points are obtained from the BM ones ({\it i.e.} the standard RSRG fixed points) by the substitution $p \to p^{\alpha}$.

There is however a subtlety concerning the finite size measure, {\it i.e.} the analogous of (\ref{fs}). As explained in the text, it is still valid provided one interprets $L$ as the "first arrival time" of the process at $u_L$. Indeed, if one considers $Z_L$ as
\begin{eqnarray}\label{fs_ctrw}
\fl Z_L = \int_{-\infty}^\infty d u_L \int_{\ell_1,\ell_2,F_1,F_2} \bar l_\Gamma E_\Gamma(F_1,\ell_1)E_\Gamma(F_2,\ell_2) \delta(L-(\ell_1+\ell_2))
\delta(u_L-(u_0+F_1-F_2)) \nonumber \\ 
\fl + \int_{-\infty}^\infty d u_L \sum_{k=1}^{\infty}  \int_{\ell_{1}, F_{1}}\bar l_\Gamma  E_\Gamma(F_{1},\ell_{1} )  
 \prod_{j=2}^{2k+1} \int_{\ell_j,F_j} P(F_{j},\ell_{j})  \int_{\ell_{2k+2}, F_{2k+2}} E_\Gamma(F_{2k+2},\ell_{2k+2} )  \delta(L- \sum_{i=1}^{2 k+2} \ell_i) \\
\fl \times \delta(u_L-(u_0+\sum_{j=1}^{2k+2} (-1)^{j+1} F_{j})) \;,
\end{eqnarray}
then one has \cite{sinaiRSRG}
\begin{eqnarray} \label{normalization_app}
&& \int_0^\infty dL e^{- p L} Z_L =  \bar l_\Gamma \frac{E_\Gamma(p)^2}{1- P_\Gamma(p)^2} = \frac{1}{p^\alpha} \;,
\end{eqnarray}
where we have used $\bar l_\Gamma=\Gamma^{2/\alpha}$ (the {\it typical} bond length) and the above forms for the fixed point (\ref{bulk_sol_ctrw}), (\ref{sol_edge_bond_app}). Therefore, $Z_L \neq 1$ for $\alpha \neq 1$ and this measure is not normalized (\ref{fs_ctrw}) if $\alpha \neq 1$. The reason for this is that if one considers a fixed time $L$, one must then convolute the finite size measure by the waiting time function $\Psi(\ell)$, which, in the language of the stochastic process corresponding to CTRW (discussed in the main text), is the probability of no jump in the time interval $[\ell,\infty [$. It is straightforward to see that its Laplace transform behaves as $\Psi(p) = A p^{\alpha-1}$ and the condition of normalization $\int_0^L e^{-pL }Z_L \Psi(p) = 1/p$ yields simply $A=1$. 

Another way to present our result for the finite size measure is to state that the correct generalization of (\ref{fs}) for models with no first moment in the bond length distribution, and statistical independence of successive bonds, reads:
\begin{eqnarray}\label{fs_ctrw2}
\fl \tilde Z_L = \int_{-\infty}^\infty d u_L \int_{\ell_1,\ell_2,F_1,F_2} \bar l_\Gamma E_\Gamma(F_1,\ell_1)E_\Gamma(F_2,\ell_2) \Psi((\ell_1+\ell_2)-L)  \\ 
\fl + \int_{-\infty}^\infty d u_L \sum_{k=1}^{\infty}  \int_{\ell_{1}, F_{1}}\bar l_\Gamma  E_\Gamma(F_{1},\ell_{1} )  
 \prod_{j=2}^{2k+1} \int_{\ell_j,F_j} P(F_{j},\ell_{j})  \int_{\ell_{2k+2}, F_{2k+2}} E_\Gamma(F_{2k+2},\ell_{2k+2} )  \Psi(\sum_{i=1}^{2 k+2} \ell_i - L) \nonumber 
\end{eqnarray}
with the "waiting time" function $\Psi$ discussed above, such that $\tilde Z_L$ is then correctly normalized to unity. This measure is clearly invariant under the RSRG procedure, up to the flow of $P_\Gamma$ and $E_\Gamma$ as described above.
This means that the $\delta$-function constraint on total bond length in the finite size measure is only possible for the BM class. It would be very interesting to study whether a starting landscape e.g. with a fixed total length and number of bonds will indeed flow, and in which sense, to this asymptotic form. At this stage our main argument is based on (i) invariance of the form (\ref{fs_ctrw2}) under RSRG (ii) the CTRW interpretation given above. More generally, convergence to finite size measures has not, to our knowledge, been studied and is a fascinating subject. This however is beyond the scope of this paper.

\section{RSRG method for the extremum on the edge bond}\label{appendix_edge_max}

Let us now call $E_\Gamma(F,\ell,u,x)$ the joint probability that the edge bond has $F,\ell$ and a maximum at $x$ of value $u$. Then it satisfies:
\begin{eqnarray}
&& \fl  \partial_\Gamma E_\Gamma(F,\ell,u,x) = - P_\Gamma(0) E_\Gamma(F,\ell,u,x) \\
&& \fl  + \int_{F_1, \zeta_3}\int_{\ell_1+\ell_2+\ell_3=\ell}
E_\Gamma(F_1,\ell_1,u_1,x_1) P_\Gamma(0,\ell_2)
P_\Gamma(\zeta_3,\ell_3) \delta(F-(F_1+\zeta_3)) \\
&& \fl \times \big[ 
\theta(u_1-(\Gamma-F_1)) \delta(u-u_1)\delta(x-x_1) + \theta(\Gamma-F_1-u_1) \delta(u-(\Gamma-F_1))
\delta(x-(l_1+l_2)) \big] \nonumber \;.
\end{eqnarray}
This is equivalent to:
\begin{eqnarray}
&& \fl \partial_\Gamma E_\Gamma(F,\ell,u,x) = - P_\Gamma(0)
 E_\Gamma(F,\ell,u,x)\\
&& \fl + \int_{\max(0,\Gamma-u)}^F  dF_1  \int_{\ell_1+\ell_2+\ell_3=\ell}
E_\Gamma(F_1,\ell_1,u,x) P_\Gamma(0,\ell_2) P_\Gamma(F-F_1,\ell_3) \nonumber 
\\
&& \fl + \theta(\Gamma-u) 
\int_{0}^u du_1  \int_0^x d\ell_1 \int_{0}^{\ell_1} dx_1 
E_\Gamma(\Gamma - u,\ell_1,u_1,x_1) P_\Gamma(0,x-\ell_1)
 P_\Gamma(F+u-\Gamma,\ell-x) \;. 
 \nonumber
\end{eqnarray}
In Laplace variables w.r.t $\ell$ and $x$ we get:
\begin{eqnarray}
&& \fl \partial_\Gamma E_\Gamma(F,p,u,q) = - P_\Gamma(0) E_\Gamma(F,p,u,q) + \int_{m(0,\Gamma-u)}^F  dF_1  
E_\Gamma(F_1,p,u,q) P_\Gamma(0,p) P_\Gamma(F-F_1,p) \nonumber 
\\
&& \fl + \theta(\Gamma-u) 
\int_{0}^u du_1 
E_\Gamma(\Gamma - u,p+q,u_1,q=0) P_\Gamma(0,p+q) P_\Gamma(F+u-\Gamma,p) \;.
 \nonumber
\end{eqnarray}

From the Markov property of the BM the solution of this equation must take the form:
\begin{eqnarray}
&& E_\Gamma(F,\ell,u,x) = A_\Gamma(F,u,x) B_\Gamma(u+F,\ell-x) \;, \\
&& E_\Gamma(F,p,u,q) = A_\Gamma(F,u,p+q) B_\Gamma(u+F,p) \;,
\end{eqnarray}
where $A$ is the sum over paths starting at $0,0$ ending at $x,u$ constrained to remain on interval $[-F,u]$ and
with no return of more than $\Gamma$, while $B$ are paths starting at $x,u$ ending at $\ell,-F$ constrained to remain on the interval $[-F,u]$
with no return of more than $\Gamma$. For $\Gamma>u$ the return constraint does not play a role for $A$, for $\Gamma>u+F$ it does not play a role for $B$. Whenever return constraint does not play a role one has:
\begin{eqnarray}
&& A_\Gamma(F,u,p) = \frac{\sinh(\sqrt{p} F)}{\sinh(\sqrt{p} (u+F))} \;, \\
&& B_\Gamma(u+F,p)= \frac{\sqrt{p}}{\sinh(\sqrt{p} (u+F))} \;,
\end{eqnarray}
{\it i.e.} independent of $\Gamma$.
For smaller $\Gamma$ one must solve the full equation above with initial condition 
$E_\Gamma(F=\Gamma,\ell,u,x) = \delta(u)\delta(x) E_\Gamma(F=\Gamma,\ell)$. We will not attempt this here as we are mostly interested in the large $\Gamma$ limit. 

Hence our final result for the solution at large $\Gamma$ reads:
\begin{eqnarray}
&& E_\Gamma(F,p,u,q) = \frac{1}{\Gamma} \frac{ \sinh(\sqrt{p+q}
    F)}{\sinh(\sqrt{p+q} (u+F)) } \frac{\sqrt{p}}{\sinh(\sqrt{p}
    (u+F))} \;.
\end{eqnarray}
One can check that
\begin{eqnarray}
&& \int_0^\infty du E_\Gamma(F,p,u,0) = \frac{1}{\Gamma} e^{-\sqrt{p}
    F} \;,
\end{eqnarray}
which is the correct result. 

We can now compute $E_\Gamma(F,\ell,u,x)$ by performing a double inverse Laplace transform. One has indeed
\begin{eqnarray}
LT^{-1}_{q \to x}
\left(\frac{\sinh{(\sqrt{p+q}F})}{\sinh{(\sqrt{p+q}(u+F)})}\right) =  2\pi
\sum_{n=0}^\infty (-1)^{n+1} n  \frac{\sin(\frac{\pi F}{u+F}
  n)}{(u+F)^2} e^{- p x - \frac{\pi^2 n^2}{(u+F)^2} x} \;.
\end{eqnarray}
Note that this yields the identity (setting $p=0$ and
taking the Laplace 
transform wrt to $x$ of both sides)
\begin{eqnarray}\label{identite_1}
2 \pi \sum_{n=0}^\infty (-1)^{n+1} n  \frac{\sin(\frac{\pi F}{u+F}
  n)}{(u+F)^2} \frac{1}{q + \frac{n^2 \pi^2}{(u+F)^2}} =
 \frac{\sinh{(\sqrt{q}F})}{\sinh{(\sqrt{q}(u+F)})} \;,
\end{eqnarray}
which we also checked with Mathematica.

Next we have
\begin{eqnarray}\label{inverselaplace_1}
LT^{-1}_{p \to y} \left(\frac{\sqrt{p}e^{-px}}{\sinh{(\sqrt{p}(u+F))}}
\right) = 2 \pi^2 \frac{1}{(u+F)^3}\sum_{m=0}^\infty (-1)^{m+1} m^2
e^{-\frac{\pi^2 m^2}{(u+F)^2}(y-x)} \;. 
\end{eqnarray}
Note that this yields trivially the identity (setting $x=0$ and
taking the Laplace 
transform wrt to $y$ of both sides)
\begin{eqnarray}\label{identite_2}
2 \pi^2 \frac{1}{(u+F)^3}\sum_{m=0}^\infty (-1)^{m+1}\frac{m^2}{p +
\frac{m^2 \pi^2}{(u+F)^2}} = \frac{\sqrt{p}}{\sinh{(\sqrt{p}(u+F))}} \;,
\end{eqnarray}
which we also checked with Mathematica. Note that the sum in the left
hand side has to be understood as 
\begin{eqnarray}\label{inverselaplace_2}
\fl 2 \pi^2 \frac{1}{(u+F)^3}\sum_{m=0}^\infty (-1)^{m+1}\frac{m^2}{p +
\frac{m^2 \pi^2}{(u+F)^2}} = \lim_{\alpha \to -1^+} 2 \pi^2 \frac{1}{(u+F)^3}\sum_{m=0}^\infty \alpha^{m+1}\frac{m^2}{p +
\frac{m^2 \pi^2}{(u+F)^2}} \;.
\end{eqnarray}

Finally, combining the both Laplace inversion in Eq. (\ref{inverselaplace_1}, \ref{inverselaplace_2}) one obtains the formula (\ref{E_direct}) given in the text.

\section{Details about reflected BM, excursions and meanders}\label{appendix_minmax}

Here we give the details about the computation of the joint
distribution $P_L(u_m,x_m,u,x,u_L)$ of the global minimum $u$ and
maximum $u_m$ and their positions $x$ and $x_m$. We start with the
expression given in the text in Eq. (\ref{min_max}):
\begin{eqnarray}\label{min_max_appendix}
&& P_L(u_m,x_m,u,x,u_L) = \theta(x_m-x) \int_{u}^{u_L} du_2
  \int_{x_m}^L dx_2 P_L(u,x,u_m,x_m,u_2,x_2,u_L) \\ 
&& + \theta(x-x_m)\int_{u}^{0} du_1  \int_{0}^{x_m} dx_1
P_L(u_1,x_1,u_m,x_m,u,x,u_L)  \;.
\end{eqnarray}

Let us first focus on the first term. The integral over the space
variable $x_2$ can be done by noticing that it can be written,
formally, as
\begin{eqnarray}
&& \fl \int_{u}^{u_L} du_2
  \int_{x_m}^L dx_2 P_L(u,x,u_m,x_m,u_2,x_2,u_L) = \int_{u}^{u_L} du_2
  \int_{x_m}^L dx_2 F(L-x_2) G(x_2-x_m) \nonumber \\
&&= \int_{u}^{u_L} d u_2
  \int_{0}^{L-x_m} dy_2 F(L-x_m-y_2) G(y_2) \;, 
\end{eqnarray}
where the functions $F, G$ can be read straightforwardly on
Eq. (\ref{full_expr}). The integral over $y_2$ can be performed by
taking the Laplace 
transform with respect to $\tilde L = L-x_m$. This yields
\begin{eqnarray}
&&\int_{0}^{L} dy e^{-p \tilde L} \int_{u}^{u_L} d u_2
  \int_{0}^{\tilde L} dy_2 F(\tilde L-y_2) G(y_2)  \\
&& =  \sum_{n_1,m_1}^\infty  4 \pi^3 n_1 m_1^2 
  (-1)^{n_1+m_1} \frac{\sin(\frac{\pi u_m}{u_m-u} n_1)}{(u_m-u)^5}
  e^{- \frac{\pi^2}{(u_m-u)^2} (n_1^2 x + m_1^2 (x_m-x))} 
\\   
&& \fl  \times \int_{u}^{u_L} du_2 \sum_{n_2,m_2=0}^{\infty}(4 \pi^3)
  (-1)^{n_2+m_2} n_2 m_2^2 
  \frac{\sin(\frac{\pi (u_m-u_L)}{u_m-u_2} n_2)}{(u_m-u_2)^5}
  \frac{1}{p + \frac{n_2^2\pi^2}{(u_m-u_2)^2}} \frac{1}{p +
  \frac{m_2^2\pi^2}{(u_m-u_2)^2}} \;.  
\end{eqnarray}
Now one can use the identities in Eq. (\ref{identite_1},
\ref{identite_2}) to perform the sums over $n_1, m_1$ to obtain
\begin{eqnarray}
&&\int_{0}^{L} dy e^{-p \tilde L} \int_{u}^{u_L} d u_2
  \int_{0}^{\tilde L} dy_2 F(\tilde L-y_2) G(y_2) \\
&& =  \sum_{n_1,m_1}^\infty  4 \pi^3 n_1 m_1^2 
  (-1)^{n_1+m_1} \frac{\sin(\frac{\pi u_m}{u_m-u} n_1)}{(u_m-u)^5}
  e^{- \frac{\pi^2}{(u_m-u_1)^2} (n_1^2 x + m_1^2 (x_m-x))} \nonumber \\
&&\times \int_{u}^{u_L} du_2 \frac{\sqrt{p}
  \sinh{(\sqrt{p}(u_m-u_L))}}{\sinh{(\sqrt{p}(u_m-u_2))}^2} \\
&& \fl =  \sum_{n_1,m_1}^\infty  4 \pi^3 n_1 m_1^2 
  (-1)^{n_1+m_1} \frac{\sin(\frac{\pi u_m}{u_m-u} n_1)}{(u_m-u)^5}
  e^{- \frac{\pi^2}{(u_m-u)^2} (n_1^2 x + m_1^2 (x_m-x))}\frac{\sinh{(\sqrt{p}(u_L-u))}}{ \sinh{(\sqrt{p}(u_m-u))}} \;.
\end{eqnarray}
One can then inverse the Laplace transform (see
Eq. (\ref{inverselaplace_1})) to obtain
\begin{eqnarray}
&&\int_{u}^{u_L} du_2
  \int_{x_m}^L dx_2 P_L(u,x,u_m,x_m,u_2,x_2,u_L) \\
&& = \sum_{n_1,m_1}^\infty  4 \pi^3 n_1 m_1^2 
  (-1)^{n_1+m_1} \frac{\sin(\frac{\pi u_m}{u_m-u} n_1)}{(u_m-u)^5}
  e^{- \frac{\pi^2}{(u_m-u)^2} (n_1^2 x + m_1^2 (x_m-x))} 
\\   
&& \times  \sum_{n_2=0}^\infty 2 \pi (-1)^{n_2+1} n_2  \frac{\sin(\frac{\pi
  (u_L-u)}{u_m-u} n_2)}{(u_m-u)^2} e^{-
  \frac{n_2^2\pi^2}{(u_m-u)^2}(L-x_m)} \;.
\end{eqnarray}

The second term in Eq. (\ref{min_max}) can be computed in a similar way to get
finally the expression given in Eq. (\ref{min_max_explicit}).

\section{Marginal distribution of $x_m$ for the Bessel bridge : link with the result of Pitman and Yor}\label{appendix_pityor}

We start from the joint distribution $P^{\rm Bessel \, bridge}_L(u_m,x_m)$ given in the text in Eq. (\ref{joint_bessel}). We remind it here:
\begin{eqnarray}\label{joint_bessel_app}
\fl P^{\rm Bessel \, bridge}_L(u_m,x_m) = \frac{8
  L^{\nu+1}}{\Gamma(1+\nu) u_m^{5+2\nu}} \sum_{n,m} \frac{
  j_{\nu,n}^{\nu+1} j_{\nu,m}^{\nu+1}    }{J_{\nu+1}(j_{\nu,n})
  J_{\nu+1}(j_{\nu,m})} e^{- \frac{j_{\nu,n}^2 x_m}{u_m^2}}  e^{-
  \frac{j_{\nu,m}^2 (L-x_m)}{u_m^2}} \;.
\end{eqnarray}
The marginal distribution of the maximum $u_m$ is obtained after
integration over $x_m$. It reads  
\begin{eqnarray}\label{marginal_us}
&&{\rm P}^{\rm Bessel \, bridge}_L(u_m) =  \frac{1}{\sqrt{L}} \tilde
  {\rm P}^{\rm Bessel \, bridge}\left(\frac{u_m}{\sqrt{L}}\right)
  \nonumber \\
&&\tilde {\rm P}^{\rm Bessel \, bridge}(x) = \frac{8}{\Gamma(1+\nu) x^{3+2\nu}} \sum_{n,m} \frac{   j_{\nu,n}^{\nu+1} j_{\nu,m}^{\nu+1}    }{J_{\nu+1}(j_{\nu,n})   J_{\nu+1}(j_{\nu,m})}\frac{   e^{- \frac{j_{\nu,n}^2}{x^2}} -    e^{- \frac{j_{\nu,m}^2}{x^2         }}} {j_{\nu,m}^2-j_{\nu,n}^2} \;,
\end{eqnarray}
while Pitman and Yor found in Ref. \cite{pitman_yor} (note that the
value of the diffusion constant used in their paper is smaller than
ours by a factor of $2$) 
\begin{eqnarray}\label{marginal_pityor}
 \tilde {\rm P}^{\rm Bessel \, bridge}(x) = \frac{d}{dx} \Bigg(
 \frac{4}{\Gamma(\nu+1) x^{2\nu+2}} \sum_{n=1}^{\infty}
 \frac{j_{\nu,n}^{2\nu}}{J^2_{\nu+1}(j_{\nu,n})}
 \exp{\left(-\frac{j_{\nu,n}^2}{x^2} \right)}  \Bigg) \;.
\end{eqnarray}
We want to show that these two expressions in Eq. (\ref{marginal_us}) and Eq. (\ref{marginal_pityor}) are identical and therefore we want to show the identity
\begin{eqnarray}\label{identity}
&&\frac{2}{x^{3+2\nu}} \sum_{n=1,m=1}^\infty \frac{   j_{\nu,n}^{\nu+1} j_{\nu,m}^{\nu+1}    }{J_{\nu+1}(j_{\nu,n})   J_{\nu+1}(j_{\nu,m})}\frac{   e^{- \frac{j_{\nu,n}^2}{x^2}} -    e^{- \frac{j_{\nu,m}^2}{x^2         }}} {j_{\nu,m}^2-j_{\nu,n}^2}  \\
&& = \frac{2}{x^{2\nu+5}} \sum_{n=1}^{\infty} \frac{j_{\nu,n}^{2\nu+2}}{J^2_{\nu+1}(j_{\nu,n})} \exp{\left(-\frac{j_{\nu,n}^2}{x^2} \right)} - \frac{2\nu+2}{ x^{2\nu+3}} \sum_{n=1}^{\infty} \frac{j_{\nu,n}^{2\nu}}{J^2_{\nu+1}(j_{\nu,n})} \exp{\left(-\frac{j_{\nu,n}^2}{x^2} \right)} \;.
\end{eqnarray}
Now in the double sum over $n,m$ in the left hand side of
Eq. (\ref{identity}) we separate out the $m=n$ term since it exactly
cancels the first term in the right hand side of
Eq. (\ref{identity}). Finally using a formula obtained by Pitman and
Yor in Ref. \cite{pitman_yor} (see their formula (125), notice however
that there is a misprint in their formula (124)) one obtains 
\begin{eqnarray}
\fl 2 \sum_{n\neq m=1}^\infty \frac{   j_{\nu,n}^{\nu+1}
  j_{\nu,m}^{\nu+1}    }{J_{\nu+1}(j_{\nu,n})
  J_{\nu+1}(j_{\nu,m})}\frac{   e^{- \frac{j_{\nu,n}^2}{x^2}} -
  e^{- \frac{j_{\nu,m}^2}{x^2         }}} {j_{\nu,m}^2-j_{\nu,n}^2} =
- 2(\nu +1) \sum_{n=1}^{\infty}
\frac{j_{\nu,n}^{2\nu}}{J^2_{\nu+1}(j_{\nu,n})}
\exp{\left(-\frac{j_{\nu,n}^2}{x^2} \right)} \;,
\end{eqnarray}
which shows finally the identity in Eq. (\ref{identity}). 

%\begin{thebibliography}{32}

%\end{thebibliography}

\end{document}